\documentclass[
 reprint,
superscriptaddress,
showkeys,
 amsmath,amssymb,
 aps,
prb,
floatfix
]{revtex4-2}

\usepackage{graphicx}
\usepackage{dcolumn}
\usepackage{bm}

\begin{document}

\preprint{APS/123-QED}

\title{Ferroaxial and nematic transitions in the charge density wave phase of 1T-TiSe$_2$}

\author{Sarah Edwards}
    \email{sarahe42@uw.edu}
    \affiliation{Department of Physics, University of Washington, Seattle, Washington 98195, USA}
\author{Elliott Rosenberg}%
    \affiliation{Department of Physics, University of Washington, Seattle, Washington 98195, USA}
    \affiliation{Department of Physics, Lehigh University, Bethlehem, PA 18015}%

\author{Ilaria Maccari}

\affiliation{Institute for Theoretical Physics, ETH Zürich, 8093 Zürich, Switzerland}

\author{Jiaqin Wen}
\affiliation{Department of Physics, University of Washington, Seattle, Washington 98195, USA}

\author{Chaowei Hu}

\affiliation{Department of Physics, University of Washington, Seattle, Washington 98195, USA}

\author{Xiaodong Xu}

\affiliation{Department of Physics, University of Washington, Seattle, Washington 98195, USA}
\affiliation{Department of Materials Science and Engineering, University of Washington, Seattle, Washington 98195, USA}

\author{Jong-Woo Kim}

\affiliation{Advanced Photon Source, Argonne National Laboratory, Lemont, Illinois 60439, USA}

\author{Philip J. Ryan}

\affiliation{Advanced Photon Source, Argonne National Laboratory, Lemont, Illinois 60439, USA}

\author{Rafael M. Fernandes}
\affiliation{Department of Physics, The Grainger College of Engineering, University of Illinois Urbana-Champaign, Urbana, Illinois 61801, USA}
\affiliation{Anthony J. Leggett Institute for Condensed Matter Theory, 
The Grainger College of Engineering, University of Illinois Urbana-Champaign, Urbana, Illinois 61801, USA}

\author{Fernando de Juan}

\affiliation{Donostia International Physics Center, 20018 Donostia-San Sebastian, Spain}
\affiliation{IKERBASQUE, Basque Foundation for Science, 48013 Bilbao, Spain}

\author{Maria N. Gastiasoro}
\email{maria.ngastiasoro@dipc.org}
\affiliation{Donostia International Physics Center, 20018 Donostia-San Sebastian, Spain}

    \author{Jiun-Haw Chu}
\email{jhchu@uw.edu}
\affiliation{Department of Physics, University of Washington, Seattle, Washington 98195, USA}

\date{\today}

\begin{abstract}
Charge density waves (CDWs) with multi-component order parameters can break unexpected symmetries through the interplay of nearly degenerate instabilities. In the widely investigated material 1T-TiSe$_2$, a central question is whether the observed CDW has a chiral character, which would manifest as the spontaneous breaking of mirror and inversion symmetries. Previous experiments have reported conflicting results about the broken symmetries in the CDW phase of 1T-TiSe$_2$. Here, we resolve this controversy by identifying the bulk broken symmetry as ferroaxial, corresponding to the breaking of vertical mirrors while preserving inversion symmetry. Using symmetry-resolved elastoresistivity, we detect the spontaneous emergence of intrinsic off-diagonal elastoresistivity coefficients that satisfy an antisymmetric relation ($m_{xx-yy,xy} \approx -m_{xy,xx-yy}$), providing an unambiguous bulk transport signature of a macroscopic electric toroidal moment. Simultaneous elastocaloric measurements reveal that the onset of ferroaxial order occurs just below the CDW transition. As the temperature is lowered further, a diverging nematic susceptibility signals a distinct rotational symmetry-breaking instability inside the ferroaxial CDW state. Our findings demonstrate that the proposed ``chiral'' CDW in 1T-TiSe$_2$ is actually a centrosymmetric ferroaxial state, reconciling previous surface-sensitive observations with bulk symmetry constraints.
\end{abstract}

\maketitle
\begin{figure*}
\centering
\includegraphics[width=0.9\textwidth]{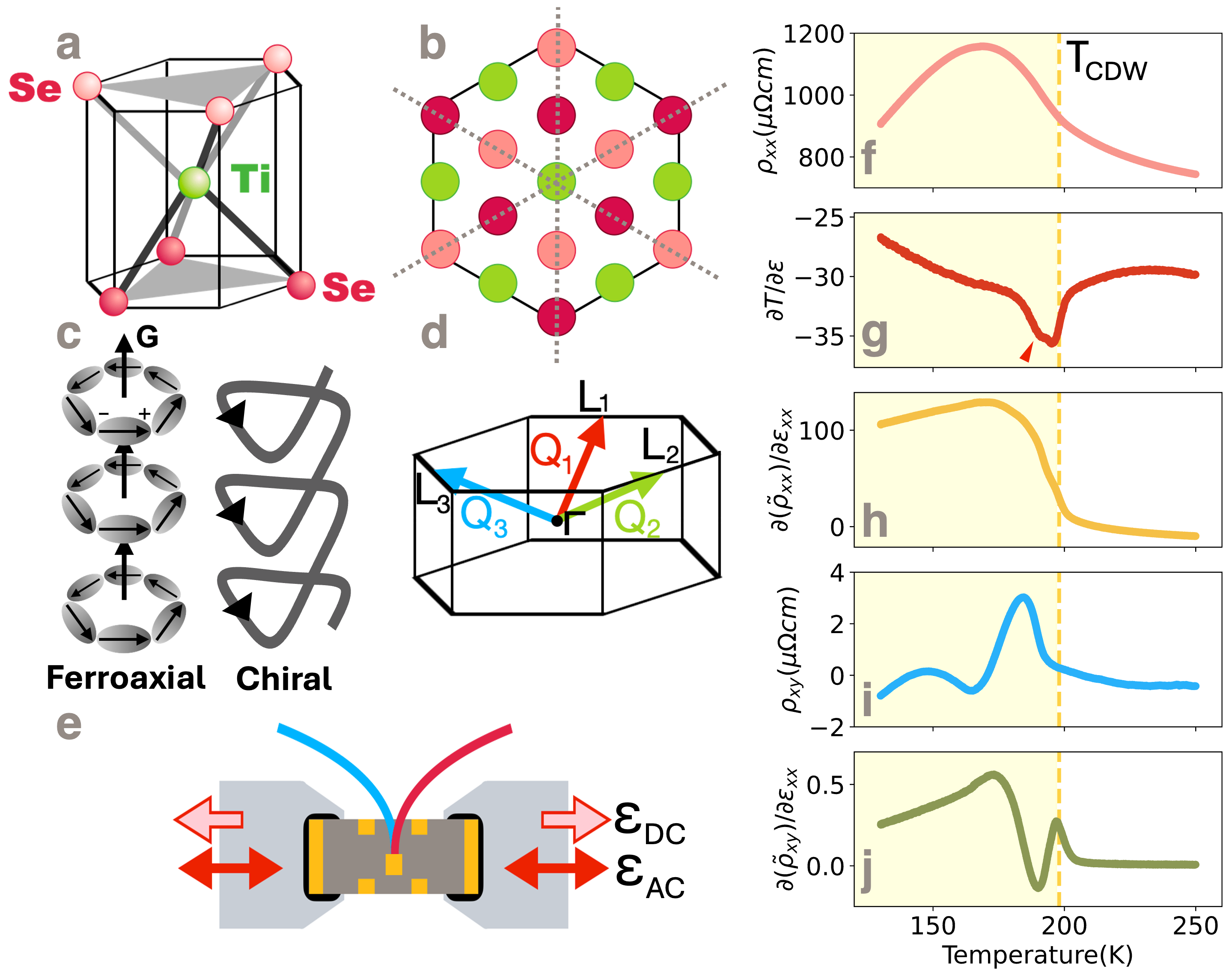}
\caption{\textbf{Experimental signatures of symmetry breaking in 1T-TiSe$_2$.} 
\textbf{a.} Crystal structure of 1T-TiSe$_2$, showing the trigonal unit cell with point group D$_{3d}$. 
\textbf{b.} A top-down view of a single quasi-2D layer. Dashed lines indicate the dihedral (vertical) mirror planes 
\textbf{c.} Schematics distinguishing ferroaxial order (characterized by an electric toroidal dipole moment $\mathbf{G}$) from chiral order (proposed for this material as a helical charge ordering \cite{Ishioka2010}). 
\textbf{d.} 3D Brillouin zone showing the three charge density wave vectors ($Q_1, Q_2, Q_3$) connecting $\Gamma$ to the $L$ points. 
\textbf{e.} Schematics of the AC strain experimental setup. A bar-shaped single crystal is suspended between two piezoelectric stacks. A type-E thermocouple (Constantan/Chromel) and electrical contacts (gold) allow for simultaneous measurements of temperature oscillations (elastocaloric) and resistivity (elastoresistivity) while combined DC strain offset ($\varepsilon_{DC}$) and AC strain oscillation ($\varepsilon_{AC}$) are applied. 
\textbf{f--j.} Transport and thermodynamic responses measured at zero DC strain ($\varepsilon_{DC} = 0$). 
\textbf{f.} Longitudinal resistivity ($\rho_{xx}$) as a function of temperature, showing the characteristic broad hump that onsets at $T_{CDW}$ (vertical dashed line). 
\textbf{g.} Elastocaloric coefficient $\partial T / \partial \varepsilon$ (red curve). Note the fine structure (a shoulder) near the transition, indicative of the two distinct phase boundaries discussed in the text. The shoulder is marked with a red arrow.
\textbf{h.} Longitudinal elastoresistivity $\partial\tilde{\rho}_{xx} / \partial \varepsilon_{xx}$ (yellow curve), exhibiting a giant step-like increase at the transition. 
\textbf{i.} Transverse resistivity ($\rho_{xy}$) as a function of temperature measured in \textit{zero magnetic field}. The spontaneous nonzero value indicates symmetry breaking. 
\textbf{j.} Off-diagonal elastoresistivity $\partial\tilde{\rho}_{xy} / \partial \varepsilon_{xx}$ (green curve). This coefficient is symmetry-forbidden in the high-temperature phase but turns on at $T_{CDW}$.
} 
\label{fig1}
\end{figure*}

 The charge density wave (CDW) of 1T-TiSe$_2$ has remained a subject of intense scrutiny for nearly five decades \cite{DiSalvo1976}, serving as a prototypical platform for investigating collective electronic behaviors in van der Waals transition metal dichalcogenides. While the material is known to undergo a CDW transition at $T_{CDW} \approx 200$ K into a commensurate $2\times 2\times 2$ superlattice, manifested by band folding~\cite{Watson2019a}, its driving mechanism and the precise symmetries it breaks remain hotly debated~\cite{Ishioka2010, vanWezel2011,Castellan2013, commenton, Hildebrand2018, Xu2020, Wickramaratne2022, Lin2024,Kim2024, Ueda25}. Beyond the possible realization of an excitonic insulator \cite{Cercellier2007, Kogar2017a, Bok2021, Pashov2025}, a key open question is whether the CDW breaks the rotational, mirror, and inversion symmetries of the high-temperature trigonal phase shown in Fig. \ref{fig1}a (described by the D$_{3d}$ point group). Early scanning tunneling microscopy (STM) studies have reported unequal intensities of the three charge modulation vectors ($\mathbf{Q}_i$, Fig. \ref{fig1}d), which was interpreted as a signature of chiral charge ordering~\cite{Ishioka2010}. Although subsequent x-ray diffraction, photocurrent and Raman spectroscopy studies supported this claim~\cite{Castellan2013, Xu2020,Kim2024}, several experimental and theoretical studies contested it ~\cite{commenton, Hildebrand2018,Wickramaratne2022,Ueda25}. Intriguingly, recent theoretical work proposed an alternative nematic scenario where rotational symmetry is broken \cite{MunozSegovia2025}. Resolving the symmetries broken by the CDW phase requires a macroscopic symmetry-sensitive bulk probe capable of strictly distinguishing between mirror-breaking, inversion-breaking, and rotation-breaking order parameters.

 Strictly defined, a chiral state breaks all mirror-symmetry planes as well as inversion symmetry \cite{Kusunose2024}. However, in centrosymmetric crystals like 1T-TiSe$_2$, it is possible to break mirror symmetries while preserving inversion symmetry, resulting in a state known as ferroaxial order \cite{Hlinka2016,Hayami2018,Cheong2018,Winkler2023,DayRoberts2025}. As shown in Fig. \ref{fig1}c, ferroaxial order breaks dihedral mirror planes (which we will refer to as vertical mirrors from here on, as they correspond to the three mirror planes, $\sigma_d$, that contain the principal rotation axis shown in Fig. \ref{fig1}b) while preserving inversion symmetry and (if present) the horizontal mirror plane. The order parameter is characterized by a macroscopic electric toroidal dipolar moment, $\mathbf{G}  = G \hat z$, representing a finite ``rotation'' of the electronic density \cite{Hlinka2016}. In contrast, chiral order is characterized by a scalar order parameter corresponding to an electric toroidal monopole \cite{Hayami2018,Kusunose2024,Huang2026}. 
Ferroaxiality was earlier recognized in oxide antiferromagnets \cite{Johnson2011, Johnson2012,Hearmon2012,Zeng2025}, and has recently been identified in chalcogenide CDW systems \cite{Ren2024, Singh2025}. 
Detecting ferroaxial order can be notoriously difficult. Because it preserves both time-reversal and inversion symmetries, it is ``dark'' to standard linear or non-linear electric and magnetic susceptibilities due to the lack of a clear conjugate field to the electric toroidal moment \cite{Hlinka2016}. However, symmetry analysis predicts a unique coupling of $G$ to the combination of shear and anisotropic strains. Just as a finite magnetization generates an antisymmetric off-diagonal resistivity, the formation of a ferroaxial order parameter activates specific off-diagonal elastoresistivity coefficients, as it enables a pure shear strain $\varepsilon_{xy}$ to generate an anisotropic resistivity response $(\rho_{xx}-\rho_{yy})$. This cross-coupling is symmetry-forbidden in the high-temperature trigonal or nematic phases, making it a definitive ``smoking gun'' for the presence of ferroaxial order \cite{DayRoberts2025}.

In this paper, we utilize complementary strain techniques to probe the broken symmetries of the CDW phase in 1T-TiSe$_2$. We report the discovery of a non-zero off-diagonal elastoresistivity response that onsets below $T_{CDW}$. Remarkably, the two cross-coefficients, $m_{12} \equiv m_{xy, x^2-y^2}$ and $m_{21} \equiv m_{x^2-y^2, xy}$, develop with opposite signs ($m_{12} \approx -m_{21}$). This antisymmetric relation provides unambiguous thermodynamic evidence of a bulk ferroaxial order, distinguishing it from nematic and chiral orders, and ruling out extrinsic sources such as sample misalignment. Furthermore, we observe that the nematic susceptibility, measured by the diagonal elastoresistivity coefficient $m_{11} = m_{x^2 - y^2, x^2-y^2}$, while sub-dominant at $T_{CDW}$, diverges rapidly at lower temperatures, signaling a distinct nematic instability deep within the ferroaxial phase. Complemented by elastocaloric measurements that map the true thermodynamic phase boundaries as a function of temperature and strain, our results establish a clear symmetry-breaking hierarchy: the primary CDW instability is followed by a ferroaxial transition  (whose order parameter transforms as the $A_{2g}$ irreducible representation) , which acts as the parent state for secondary nematic transition  (whose order parameter transforms as the $E_{g}$ irreducible representation) . Our results resolve the long-standing debate over the ``chiral'' nature of 1T-TiSe$_2$ and demonstrate symmetry-resolved elastoresistivity as a powerful probe for hidden centrosymmetric orders.

Figure \ref{fig1} shows the temperature evolution of several experimental quantities measured simultaneously as a sample of 1T-TiSe$_2$ is cooled from room temperature to 125 K. Figure \ref{fig1}e illustrates the experimental configuration. Key here is the use of a well established technique of employing a small oscillating AC strain superimposed on a constant DC offset to induce time-dependent changes in resistance and temperature quasi-adiabatically, enabling fast measurement of strain derivatives such as elastoresistivity and elastocaloric coefficients \cite{Ikeda2019, Hristov2018}. Panels f-j of Figure \ref{fig1} show, respectively, longitudinal resistivity ($\rho_{xx}$), elastocaloric coefficient ($\partial T/\partial \varepsilon_{xx}$), longitudinal elastoresistivity ($\partial\tilde{\rho}_{xx}/\partial\varepsilon_{xx}$), transverse resistivity ($\rho_{xy}$), and the off-diagonal elastoresistivity ($\partial\tilde{\rho}_{xy}/\partial\varepsilon_{xx}$) measured at zero DC strain. Here, we define the normalized resistivity change as $\tilde{\rho}_{ij} = \Delta\rho_{ij}/\rho_0$, where $\rho_0$ is the zero-strain value of the longitudinal resistivity. This simultaneous measurement is essential, as it allows us to strictly correlate transport anomalies with thermodynamic phase boundaries measured on the exact same sample volume.

As shown in Figure \ref{fig1}f, the zero-strain resistivity $\rho_{xx}$ exhibits a characteristic broad hump, the onset of which (defined as a sharp peak in the second derivative of resistivity with respect to temperature) has been used in many prior works to identify the CDW transition \cite{Moulding2022, Watson2019, Campbell2019, Li2022, Wang2025}. We have marked the CDW onset by a yellow dashed line on each plot. Thermodynamically, this transition is captured by the elastocaloric coefficient (Figure \ref{fig1}g), which measures the strain derivative of the entropy and is proportional to the heat capacity anomaly near a phase transition~\cite{Ikeda2019}. The CDW onset as found from resistivity aligns with the descending edge of a dip in $\partial T/\partial \varepsilon_{xx}$. Interestingly, a distinct shoulder (marked by the red arrow) of this dip indicates that there is a second transition $\sim$7K lower than $T_{CDW}$. As will be discussed later, this second transition is not due to sample inhomogeneity \textemdash rather, it is a robust feature of 1T-TiSe$_2$. Coincident with $T_{CDW}$, the longitudinal elastoresistivity $m_{xx,xx} = \partial\tilde{\rho}_{xx}/\partial\varepsilon_{xx}$ (Figure \ref{fig1}h) begins a dramatic increase that reaches a giant magnitude of $\sim 120$, which is comparable to the nematic susceptibility observed in iron-based superconductors~\cite{Chu2012}. This massive response indicates that the electronic system becomes extremely susceptible to lattice deformations upon entering the CDW phase, a point we will return to in the context of nematic fluctuations.

The most striking feature of our data is the emergence of non-zero responses in channels that are forbidden by the symmetries of both the high-temperature phase and the widely accepted 3Q CDW phase\cite{DiSalvo1976}. In the high-symmetry $D_{3d}$ phase, both the transverse resistivity $\rho_{xy}$ and the off-diagonal elastoresistivity $m_{xy,xx} = \partial\tilde{\rho}_{xy}/\partial\varepsilon_{xx}$ are strictly constrained to be zero by the vertical mirror symmetries of the parent lattice. 
Contrary to this expectation, Figure \ref{fig1}i shows that a spontaneous static $\rho_{xy}$ develops immediately below $T_{CDW}$. Even more notably, the off-diagonal elastoresistivity $m_{xy,xx}$ (Figure \ref{fig1}j) exhibits a distinct, non-monotonic variation that peaks at the thermodynamic transition. The observation of a finite response in these off-diagonal channels provides direct evidence that the macroscopic CDW order breaks the vertical mirror symmetries of the parent lattice \cite{DayRoberts2025}. 

\begin{figure*}
\centering
\includegraphics[width=0.9\textwidth]{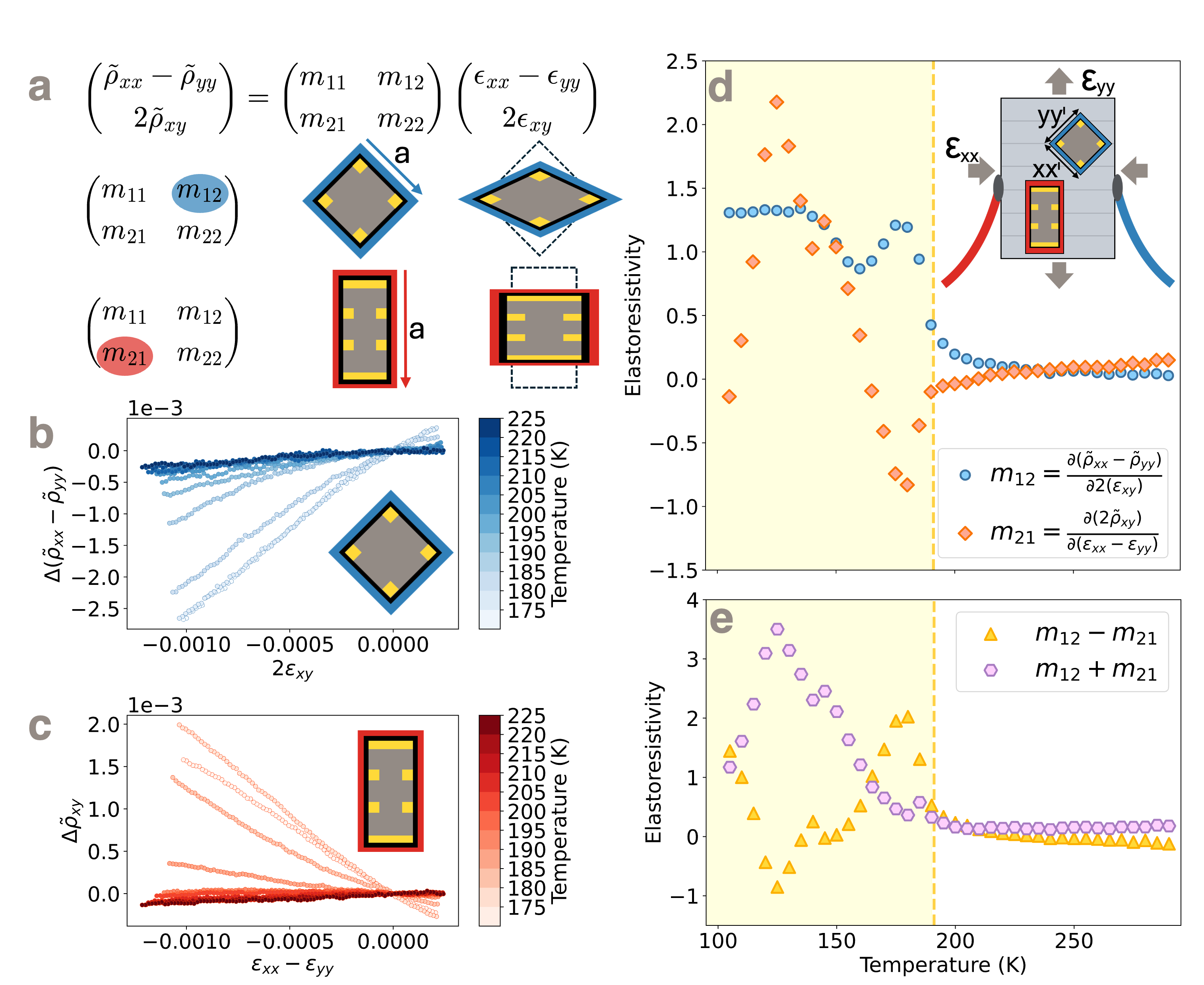}
\caption{\textbf{Ferroaxial order revealed by off-diagonal elastoresistivity.} 
\textbf{a.} The elastoresistivity tensor projected onto the subspace of the $E_g$ irreducible representation relates resistivity anisotropy components to symmetry-matched strains. In the parent $D_{3d}$ point group, diagonal elements ($m_{11}, m_{22}$) probe the nematic susceptibility, while off-diagonal elements ($m_{12}, m_{21}$) are symmetry-forbidden. Their nonzero values indicate the breaking of vertical mirror planes either by ferroaxial order ($m_{12}-m_{21}$) ~\cite{DayRoberts2025} or by the second component of the nematic order parameter ($m_{12}+m_{21}$). Diagrams illustrate the specific sample geometries used to isolate $m_{12}$ (rotated Montgomery) and $m_{21}$ (Hall bar).  
\textbf{b.} Induced change in resistivity anisotropy $\Delta(\tilde{\rho}_{xx} - \tilde{\rho}_{yy})$ versus shear strain $2\varepsilon_{xy}$, showing a linear response ($m_{12}$). 
\textbf{c.} Induced change in transverse resistivity $\Delta\tilde{\rho}_{xy}$ versus anisotropic strain $\varepsilon_{xx} - \varepsilon_{yy}$, showing a linear response ($m_{21}$). 
\textbf{d.} Temperature dependence of the extracted off-diagonal coefficients. The two signals display a striking antisymmetric relationship ($m_{12} \approx -m_{21}$) emerging near $T_{CDW}$ (vertical dashed line). \textit{Upper inset:} Experimental configuration showing the square sample rotated by $45^\circ$ relative to the poling direction of the piezo stack (applying shear strain) and the Hall bar aligned with the axis (to apply anisotropic strain). 
\textbf{e.} Symmetry decomposition of the off-diagonal response. The difference ($m_{12} - m_{21}$, yellow) tracks the ferroaxial order parameter and onsets at $T_{CDW}$. The sum ($m_{12} + m_{21}$, purple), which tracks the shear component (i.e., the second component) of the nematic order parameter, remains suppressed until $T < 170$~K, confirming the separation of the two symmetry-breaking scales.}
\label{Fig2}
\end{figure*}


However, identifying the precise symmetry of this order requires further analysis of the symmetry of elastoresistivity, in particular to rule out the misalignment effects. In the high-temperature $D_{3d}$ point group, the in-plane anisotropic resistivity components $(\tilde{\rho}_{xx} - \tilde{\rho}_{yy}, 2\tilde{\rho}_{xy})$ and strain components $(\varepsilon_{xx} - \varepsilon_{yy}, 2\varepsilon_{xy})$ both transform as the two-dimensional $E_g$ irreducible representation. Their linear coupling is described by the $2 \times 2$ elastoresistivity tensor shown in Fig.~\ref{Fig2}a, where the indices 1 and 2 correspond to the longitudinal anisotropy ($x^2-y^2$) and the shear anisotropy ($2xy$), respectively. This notation is distinct from the conventional Voigt notation, as our notation focuses only on the subspace of this two-dimensional irreducible representation. The tensor decomposes into irreducible representations of $D_{3d}$ (see Methods): the antisymmetric off-diagonal component, $m_{12} - m_{21}$, transforms as $A_{2g}$ and uniquely signifies ferroaxial order, while the symmetric combinations $(m_{11} - m_{22}, m_{12} + m_{21})$ transform as $E_g$ and serve as the nematic order parameter, which behaves as a 3-state Potts order parameter \cite{Little2020,Fernandes2020,Chakraborty2023}. The necessary condition for identifying ferroaxial order is therefore an antisymmetric off-diagonal response where $m_{12} = -m_{21}$.

To measure these coefficients, we employed DC strain measurements where samples are directly glued on the side of piezostacks (see Methods). As shown in Figure \ref{Fig2}a, for $m_{21} = \partial(2\tilde{\rho}_{xy})/\partial(\varepsilon_{xx}-\varepsilon_{yy})$, we utilized a standard Hall bar geometry aligned with the poling direction of the piezostacks. For the conjugate coefficient $m_{12} = \partial(\tilde{\rho}_{xx}-\tilde{\rho}_{yy})/\partial(2\varepsilon_{xy})$, we employed a modified Montgomery method on a square sample and the sample is rotated by $45^\circ$ with respect to the stress axis, transforming the applied uniaxial stress into a shear strain ($\varepsilon_{xy}$) in the sample's reference frame.

Figures \ref{Fig2}b and \ref{Fig2}c display the isothermal elastoresistivity response for both channels. In both geometries, the response is strictly linear, with slopes that are negligible at room temperature but increase distinctively below 200 K. The extracted temperature dependence (Fig. \ref{Fig2}d) reveals the central result of this study: $m_{12}$ and $m_{21}$ onset simultaneously near $T_{CDW}$ but with opposite signs. The coefficient $m_{21}$ (red diamonds) turns negative, while $m_{12}$ (blue circles) turns positive. This mirror-image behavior persists down to $T \approx 180$ K, where both signals exhibit a sharp non-monotonic suppression likely associated with a secondary instability. 

To disentangle the broken symmetries encoded in these observations, we decompose the off-diagonal response into its irreducible channels in Figure \ref{Fig2}e. The difference ($m_{12} - m_{21}$, yellow triangles) isolates the $A_{2g}$ component, which serves as a proxy for the ferroaxial order parameter. The sum ($m_{12} + m_{21}$, purple hexagons) corresponds to the shear component (i.e., the second component) of the nematic order parameter ($E_g$). The data reveals a striking hierarchy: the ferroaxial component onsets rapidly near $T_{CDW} \approx 200$ K, dominating immediately below the transition. The nematic component remains suppressed until significantly lower temperatures. Although the DC elastoresistivity signals in Fig.~\ref{Fig2}e do not resolve the $\approx 7$~K splitting between the CDW and ferroaxial transitions identified by our AC elastocaloric measurements (discussed below), the macroscopic symmetry-breaking hierarchy is unambiguous: the system first breaks vertical mirror symmetries while preserving rotational symmetry, with the nematic state emerging only at lower temperatures. Because $m_{12}$ and $m_{21}$ are measured on two samples, the decomposition in Fig.~\ref{Fig2}e strictly requires that both samples have comparable ferroaxial domain populations. However, this assumption is not needed to establish the $A_{2g}$ symmetry assignment: in the temperature range immediately below $T_{\mathrm{CDW}}$, where the $E_g$ order parameter has not yet condensed, the opposite signs of $m_{12}$ and $m_{21}$ can only originate from a ferroaxial contribution, independent of domain configuration.

The observations that a shear strain induces a resistivity anisotropy while an anisotropic strain induces a transverse resistivity, and that the two responses have opposite signs, is reminiscent of what happens in the anomalous Hall effect. In the latter, a non-zero magnetization $\mathbf{M}=M\hat z$ breaks time-reversal symmetry and enforces $\rho_{xy} = -\rho_{yx}$, whereas here a non-zero ferroaxial order parameter $\mathbf{G} = G\hat z$ breaks the vertical mirrors and enforces $m_{12} = -m_{21}$. This ``ferroaxial elastoresistivity'' provides unambiguous thermodynamic evidence for the existence of $\mathbf{G}$ in the CDW phase of 1T-TiSe$_2$, and effectively rules out a true chiral charge order (Fig. \ref{fig1}c right) \cite{DayRoberts2025}. Since both strain and resistivity are parity-even tensors, their linear coupling allows only for an inversion-symmetric order parameter like the ferroaxial moment, whereas parity-breaking chiral order is forbidden by symmetry from generating a linear elastoresistivity response.

\begin{figure*}
\centering
\includegraphics[width=0.9\textwidth]{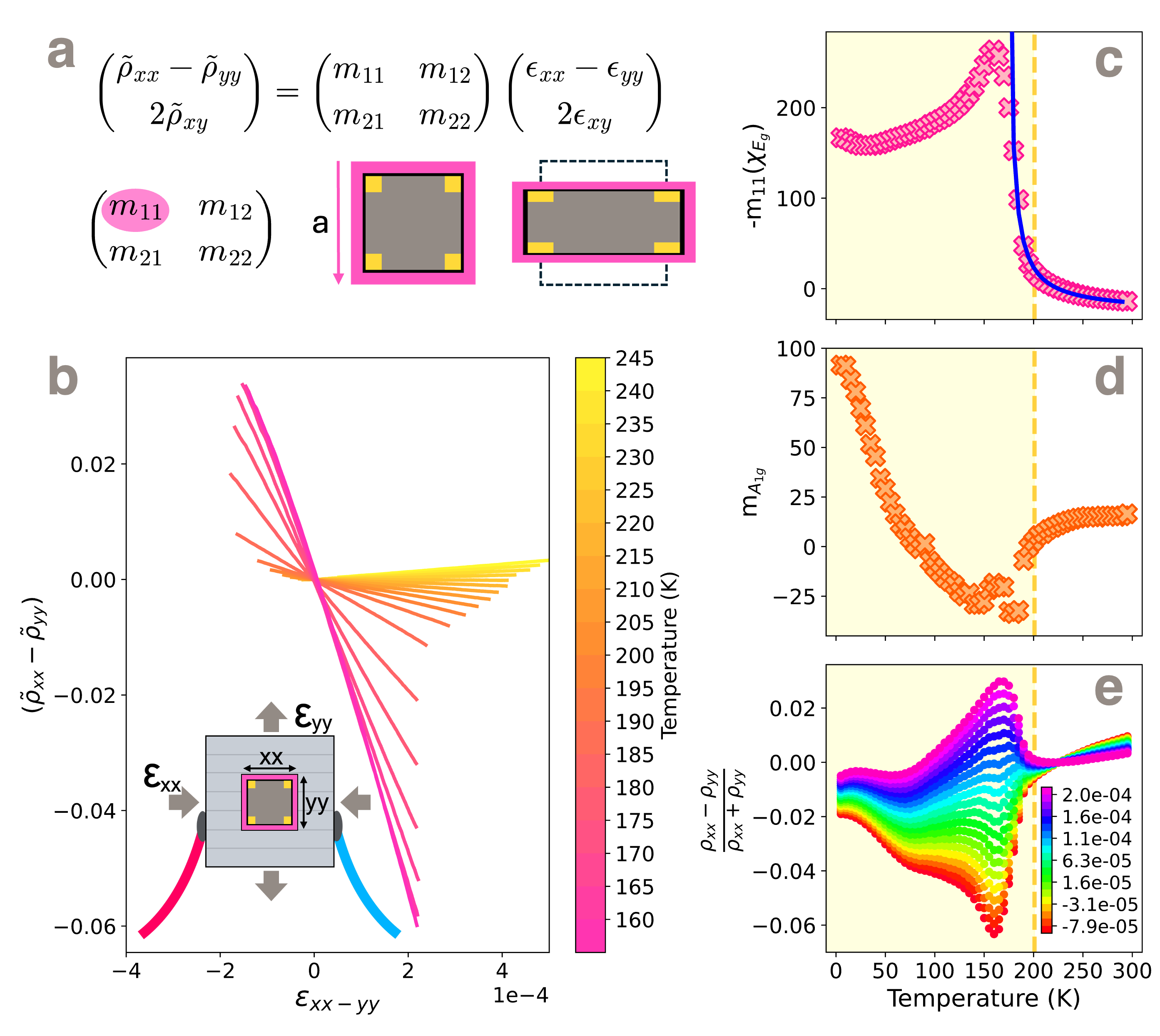}
\caption{\textbf{Diverging nematic susceptibility deep within the CDW phase.} 
\textbf{a.} \textit{Top:} The diagonal block of the elastoresistivity tensor relating the nematic order parameter (resistivity anisotropy) to the conjugate symmetry-breaking strain. \textit{Bottom:} Schematics of the experimental geometry used to measure $m_{11}$. A square single crystal is aligned with the principal strain axes of the piezoelectric stack, measuring the longitudinal resistivity anisotropy. 
\textbf{b.} Isothermal curves of the induced resistivity anisotropy $\tilde{\rho}_{xx} - \tilde{\rho}_{yy}$ versus anisotropic strain $\varepsilon_{xx} - \varepsilon_{yy}$. The response is linear over the accessible strain range, with a slope ($m_{11}$) that changes dramatically with temperature.
\textbf{c.} Temperature dependence of the diagonal element in the $E_g$ elastoresistivity block, $m_{11}$ ($\chi_{E_g}$). The response exhibits a divergent-like behavior peaking well below the CDW transition (vertical dashed line). The solid blue line represents a Curie-Weiss fit, indicating a bare nematic temperature $T^* \approx 170$~K.
\textbf{d.} Temperature dependence of the isotropic elastoresistivity coefficient $m_{A_{1g}}$. In comparison to the $E_g$ channel, this response is small and non-divergent above the nematic transition, confirming that the critical fluctuations correspond to an order parameter that breaks rotational symmetry.
\textbf{e.} Normalized resistivity anisotropy $(\rho_{xx} - \rho_{yy})/(\rho_{xx} + \rho_{yy})$ versus temperature for various fixed piezo voltage values. The color bar shows the anisotropic strain ($\varepsilon_{xx} - \varepsilon_{yy}$) values measured at $150$~K. The ``fan-like'' spread demonstrates the high tunability of the electronic anisotropy near $170$~K.} 
\label{fig3}
\end{figure*}

Turning to the diagonal elastoresistivity tensor elements, we isolated the nematic susceptibility $\chi_{E_{g}} \equiv m_{11}$ using a square Montgomery sample aligned with the piezo stack poling direction (Fig.~\ref{fig3}a), measuring the induced resistivity anisotropy $\tilde{\rho}_{xx} - \tilde{\rho}_{yy}$ as a function of the applied anisotropic strain $\varepsilon_{xx} - \varepsilon_{yy}$. Figure~\ref{fig3}b displays the isothermal resistivity anisotropy curves. The response is predominantly linear, with a slope that evolves dramatically with temperature. The extracted $\chi_{E_{g}}$ is plotted in Figure~\ref{fig3}c. In stark contrast to the off-diagonal ferroaxial signal, $\chi_{E_{g}}$ exhibits a classic Curie-Weiss-like behavior, reaching a giant magnitude exceeding 200. Crucially, the peak of $\chi_{E_g}$ occurs at $T_{nem} \approx 165$~K, significantly below $T_{CDW} \approx 200$~K. A Curie-Weiss fit to the high-temperature tail (solid blue line) yields a Weiss temperature $\theta \approx 170$~K, indicating that deep within the CDW phase, the system undergoes a distinct nematic transition.

For comparison, the isotropic elastoresistivity coefficient $m_{A_{1g}}$ (Fig.~\ref{fig3}d), defined as $\partial(\tilde{\rho}_{xx}+\tilde{\rho}_{yy})/\partial(\varepsilon_{xx}+\varepsilon_{yy})$, is an order of magnitude smaller and non-divergent above the nematic transition, displaying only a small cusp-like anomaly at $T_{CDW}$. This suggests that the giant response in Fig.~\ref{fig3}c arises from fluctuations of a rotational symmetry-breaking order parameter, and not from symmetry channel mixing. Finally, Figure \ref{fig3}e shows the resistivity anisotropy versus temperature for various fixed piezo voltages. The ability to reversibly switch the sign of the resistivity anisotropy from positive to negative by applying tensile versus compressive strain serves as strong evidence for nematicity, showing order-parameter-like behavior with a bilinear coupling to strain.


\begin{figure*}
\centering
\includegraphics[width=0.9\textwidth]{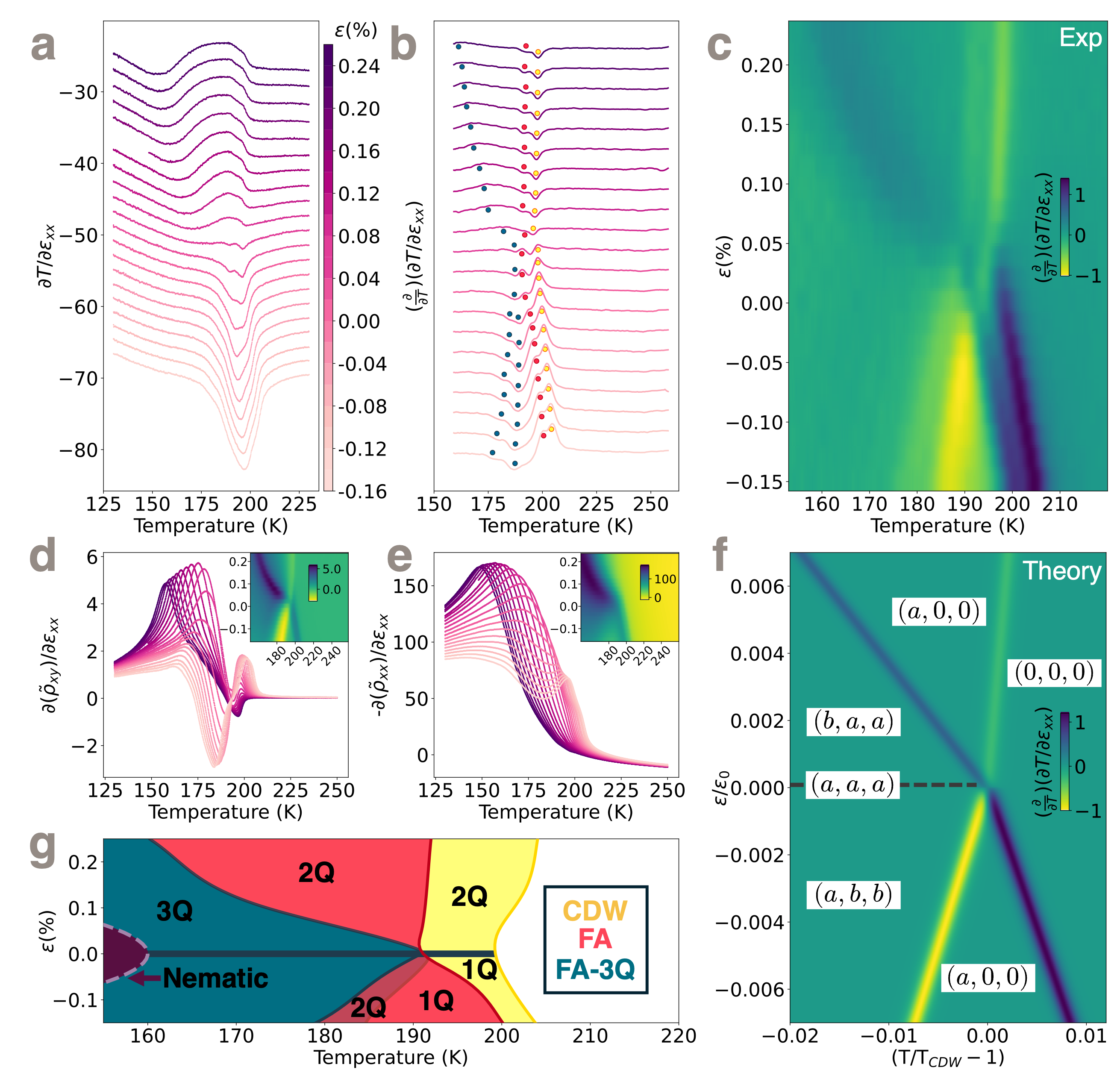}
\caption{\textbf{Strain-temperature phase diagram and the hierarchy of broken symmetries.} 
\textbf{a.} Elastocaloric coefficient ($\partial T/\partial \varepsilon_{xx}$) versus temperature for various DC strain offsets ranging from compressive (light) to tensile (dark). Curves are offset vertically for clarity.
\textbf{b.} Temperature derivative of the elastocaloric coefficient, $\frac{\partial}{\partial T}(\partial T/\partial \varepsilon_{xx})$, with curves offset vertically. The distinct peaks and anomalies, indicated with circular markers, clearly resolve the primary CDW onset, the subsequent ferroaxial transition, and the strain-split $1Q/2Q \rightarrow 3Q$ boundaries.
\textbf{c.} Color map of the temperature derivative of the elastocaloric coefficient in the strain-temperature plane. The topological structure of the bright and dark bands precisely maps the underlying phase boundaries.
\textbf{d.} Off-diagonal elastoresistivity ($\partial\tilde{\rho}_{xy}/\partial\varepsilon_{xx}$) versus temperature. \textit{Inset:} Color map of the same data, mirroring the thermodynamic boundaries established by the ECE.
\textbf{e.} Longitudinal elastoresistivity ($-\partial\tilde{\rho}_{xx}/\partial\varepsilon_{xx}$) versus temperature, serving as a proxy for the nematic susceptibility. \textit{Inset:} Color map of the longitudinal response.
\textbf{f.} Calculated temperature derivative of the ECE derived from a Landau free energy expansion. The model incorporates the bilinear coupling of the conjugate uniaxial strain field to the multi-component $3Q$ CDW order parameter. 
\textbf{g.} The comprehensive strain-temperature phase diagram constructed from the aggregated thermodynamic and transport data. The diagram illustrates the primary CDW transition, the secondary ferroaxial (FA) transition, the strain-split $1Q$ and $2Q$ phases, and the low-temperature spontaneous nematic transition within the $3Q$ phase.}
\label{fig4}
\end{figure*}

Having established the symmetry content of the elastoresistivity responses, we return to variable-strain measurements using the AC technique employed in Fig. \ref{fig1} to construct the global strain-temperature phase diagram. Figures \ref{fig4}a, \ref{fig4}d, and \ref{fig4}e display the temperature evolution of the elastocaloric coefficient ($\partial T/\partial \varepsilon_{xx}$), the off-diagonal elastoresistivity ($\partial\tilde{\rho}_{xy}/\partial\varepsilon_{xx}$) and the longitudinal elastoresistivity ($\partial\tilde{\rho}_{xx}/\partial\varepsilon_{xx}$) over a broad uniaxial strain ($\varepsilon_{xx}$) range from $-0.15\%$ (compressive) to $+0.24\%$ (tensile). These three quantities act as proxies for the thermodynamic entropy landscape, the ferroaxial order parameter, and the nematic susceptibility, respectively.

The elastocaloric coefficient exhibits a pronounced dip anomaly near $T_{\text{CDW}}$ (Fig.~\ref{fig4}a) with a strong dependence on the applied strain. Under compressive strain, two distinct dips grow in magnitude and shift to higher temperatures; under tensile strain, the anomaly evolves into a broad hump with a distinct kink indicating a splitting of the phase transition. Because each phase transition produces a step-like anomaly in the ECE (see Supplementary Information for a detailed derivation), the phase boundaries correspond to the rising and descending edges of these features. Taking the temperature derivative (Fig.~\ref{fig4}b) converts these edges into sharp peaks and dips, precisely locating the phase boundaries (highlighted by colored dots in Fig.~\ref{fig4}b). Mapping this derivative in the strain-temperature plane (Fig.~\ref{fig4}c), the CDW and subsequent ferroaxial transitions manifest as a pair of bright lines curving towards higher temperatures with increasing strain magnitude (yellow and red dots in Fig.~\ref{fig4}b). Below these, additional features (blue dots) form fan-like structures that shift to lower temperatures under increasing strain.

The topology of this thermodynamic phase diagram is strongly echoed by the transport response. The off-diagonal elastoresistivity colormap (Fig.~\ref{fig4}d, inset) mirrors the thermodynamic boundaries, and the low-temperature phase boundary aligns precisely with the maximum of the longitudinal elastoresistivity (Fig.~\ref{fig4}e), firmly establishing its connection to rotational symmetry breaking.

To interpret this phase diagram, we consider the three-component CDW order parameter $\boldsymbol{\Psi}=(\psi_1, \psi_2,\psi_3)$ transforming as the $L^-_1$ irreducible representation \cite{Bianco15,Wegner2020,Wickramaratne2022,Subedi22}, where each component corresponds to a symmetry-related $L$ point. The $\boldsymbol{\Psi}\sim\left(1,1,1\right)$ configuration gives rise to the $3Q$ state corresponding to the $2\times2\times2$ superlattice \cite{DiSalvo1976}. In the Landau free energy, the coupling of the anisotropic strain to this order parameter takes the form $\lambda_E (\varepsilon_{xx}-\varepsilon_{yy}) (\psi_1^2 - \frac{1}{2}\psi_2^2 - \frac{1}{2}\psi_3^2)$ (see Supplementary Information for a full derivation). This coupling breaks the threefold degeneracy and drives a bifurcation: for $\lambda_E>0$, compressive strain favors a $1Q$ state ($\boldsymbol{\Psi}\propto(1,0,0)$), while tensile strain favors a $2Q$ state ($\boldsymbol{\Psi}\propto(0,1,1)$) \cite{Little2020,Chakraborty2023}. Upon further cooling, the system transitions into a modified $3Q$ phase with unequal amplitudes ($\boldsymbol{\Psi}\propto(a,b,b)$, $a \neq b$). The calculated temperature derivative of the ECE from this model (Fig.~\ref{fig4}f) exhibits strong agreement with the experimental phase diagram (Fig.~\ref{fig4}c), validating our symmetry-based assignments. Direct structural confirmation is provided by \textit{in situ} elasto-X-ray diffraction at 150~K (Supplementary Fig.~S14), which shows a sevenfold enhancement of the strain-aligned $Q$-vector intensity under compressive strain, consistent with the $1Q$ assignment.

While this phenomenological model based on the $L_1^-$ CDW successfully captures the main structure of the strain-temperature phase diagram, it fails to account for the secondary phase boundary observed via ECE approximately 7~K below $T_{CDW}$. This is expected: none of the $1Q$, $2Q$, or $3Q$ configurations of the $L_1^-$ order parameter breaks the vertical mirror symmetries, and hence they cannot produce ferroaxial order. Symmetry instead requires a secondary CDW mode belonging to a distinct irreducible representation that preserves threefold rotation while breaking the vertical mirrors. Because this secondary mode is independent of the primary $L_1^-$ CDW, it is allowed to condense at a separate phase transition. As we show in Supplementary Section S5, the candidates for the secondary CDW order parameter that do not break translational symmetry and that give rise to a secondary ferroaxial order are those that transform as the $L_{2}^-$, $M_2^+$, and $A_{2}^-$ irreducible representations.  The double-peak structure resolved in our elastocaloric data, with a splitting of $\approx 7$~K, provides direct thermodynamic evidence for precisely such a sequence of two successive CDW transitions. This picture is in agreement with first-principles calculations, which consistently reveal multiple competing CDW instabilities with nearly degenerate energies \cite{Wickramaratne2022,Wegner2020,Kim2024a,Mun25}, and is analogous to the recently proposed mechanism for the ferroaxial CDW in rare-earth tritellurides \cite{Singh2025,Alekseev2024}.


Aggregating these data, we construct a comprehensive phase diagram (Fig.~\ref{fig4}g). The primary CDW onsets at $\approx 200$~K; the ferroaxial order develops at a distinct transition approximately 7~K below $T_{CDW}$. The $1Q/2Q \rightarrow 3Q$ CDW transitions define a fan-like region that widens with strain and merges with the upper transitions near zero strain. Notably, the phase diagram is asymmetric: the compressive ($1Q$) side exhibits two distinct lower-temperature boundaries, while the tensile  ($2Q$) side exhibits only one. This asymmetry
cannot be explained within the $L_1^-$ model, since the remaining vertical mirror in the $1Q$ state $(a,0,0)$ requires the simultaneous condensation of the other two components, with a direct transition into the $3Q$ state $(a,b,b)$. Hence, the intermediate transition provides additional evidence for the ferroaxial phase, which breaks the vertical mirror and allows different critical temperatures for the two components within the $2Q$ state $(a,b,0)$ (see Supplementary Information).

At lower temperatures, a 3-state Potts-nematic transition emerges at 165~K in the unstrained sample, where threefold rotational symmetry is spontaneously broken \cite{MunozSegovia2025}. The absence of a sharp elastocaloric anomaly at this temperature is consistent with an order that redistributes intensity among the three wave-vectors without substantially altering the total order parameter magnitude~\cite{Ikeda2021}. Nevertheless, we see a strong strain dependence of the magnitude of ECE, which is not predicted by the CDW only free energy (Supplementary Material, Fig. S15) but is consistent with a diverging nematic susceptibility\cite{Ikeda2021}.  Under finite strain, the fate of this Potts-nematic transition is highly sensitive to system parameters, and can evolve into 1st order, 2nd order transition or a crossover\cite{Chakraborty2023}, hence we denote this boundary with a dashed line. Taken together, the phase diagram provides a comprehensive picture of 1T-TiSe$_2$: a primary CDW instability is immediately followed by a secondary ferroaxial symmetry breaking ($A_{2g}$), setting the stage for a nematic transition ($E_g$) driven by the internal configuration of the CDW wavevectors.

This identification of the bulk symmetries broken in the CDW phase of 1T-TiSe$_2$ offers a natural resolution to conflicting experimental reports. Early X-ray and specific heat measurements identified a secondary phase transition roughly 7 K below $T_{CDW}$, initially assigned to a chiral phase \cite{Castellan2013} but later claimed to be an artifact~\cite{commenton}. Our ECE data thermodynamically validates this transition, while our elastoresistivity measurements unambiguously identify its symmetry as ferroaxial rather than chiral, which also explains why second harmonic generation is not sensitive to the transition \cite{SHG}. The key difference is that the ferroaxial phase ($A_{2g}$) preserves inversion symmetry, while the chiral phase breaks it. At the surface of a crystal or in atomically thin samples on a substrate, the inversion symmetry is extrinsically broken by the interface, and an intrinsic bulk ferroaxial order will manifest identically to a chiral state in surface-sensitive probes such as STM or photocurrent spectroscopy. Our results suggest that the previously reported chiral signatures are the surface projection of underlying bulk ferroaxial order.

Finally, these findings reframe the broader phase diagram of 1T-TiSe$_2$ with respect to other tuning knobs such as Cu-intercalation or hydrostatic pressure. It is well established that these factors suppress the CDW and stabilize a superconducting dome~\cite{Morosan2006, pressuresc}. Our discovery of distinct ferroaxial and nematic energy scales implies that the suppression of the CDW may involve a complex hierarchy of quantum phase transitions (QPTs), potentially separating a ferroaxial QPT from a 3-state Potts nematic QPT. Exploring how the fluctuations associated with rotational and mirror symmetry breaking orders interact with the formation of Cooper pairs remains an intriguing open question for future study.

\section{Methods}

\subsection{Crystal Growth}
1T-TiSe$_2$ crystals were grown using the chemical vapor transport (CVT) method, using I$_{2}$ as a transport agent under a temperature gradient of 80$^\circ$ C/m and producing large crystals several mm wide and up to 500 $\mu$m thick. Energy Dispersive X-ray (EDX) spectroscopy was used to confirm their uniform chemical composition. As has been reported in many prior works, 1T-TiSe$_2$ grown using this method tends to form with some stoichiometric defects (TiSe$_{2-\delta}$)\cite{DiSalvo1976,Huang2017, Campbell2019}. EDX measurements tell us that these samples are around $\delta = 0.14$, with only slight variations between crystals grown in the same batch.

\subsection{AC Elastoresistivity and Elastocaloric Measurements}

Data seen in Figures \ref{fig1} and \ref{fig4} are produced by single crystal samples of 1T-TiSe$_2$ cut into a hall bar geometry with the long edge parallel to the a-axis of the unit cell. In addition to the electrical contacts used to apply current and measure R$_{xx}$ and R$_{xy}$, a type E thermocouple (made up of chromel and constantan wires) was anchored to the center of the sample using silver conductive paste. A Razorbill CS100 strain cell was used to induce uniaxial strain in the samples, which were mounted between a set of two moving plates using thermally conductive epoxy (Stycast 2850 FT with catalyst 24LV). An AC voltage was applied to the outer piezoelectric stacks in the strain cell, with 2.5 Vrms amplitude and 12.5 Hz frequency, corresponding to a small periodic deformation of the sample. The frequency was chosen empirically to maximize the signal read from the thermocouple. DC voltages from 0 to 100 V were applied, using the built-in capacitive strain gauge of the cell to correlate each voltage with a near-constant strain value. The zero-point of this strain was approximated by identifying the DC voltage at which the temperature dependence of the sample resistance to data taken on the same sample before it was mounted on the strain cell most closely matched. Temperature fluctuations induced by AC strain were measured using a lock-in amplifier tuned to the frequency of the AC voltage perturbation, and the strain derivatives of longitudinal and transverse resistance were measured using the dual mode function of SRS 860 lock-in amplifiers to lock into  a sideband frequency equal to the difference of the AC current and AC strain frequencies. As has been demonstrated conclusively in past works, a voltage response proportional to the strain derivative of resistance lies at this sideband frequency \cite{Hristov2018}.

\subsection{DC Symmetry Decomposed Elastoresistivity Measurements}

Three measurements shown in this paper were conducted by gluing a single crystal sample to the sidewall of a piezoelectric stack. Two of these crystals were prepared in a square Montgomery configuration, allowing for measurement of R$_{xx}$ and R$_{yy}$,  and one in a hall bar geometry, allowing for measurement of R$_{xx}$ and R$_{xy}$. Under DC voltage bias, the piezostack expands or contracts along its long edge (the poling direction), inducing the opposite direction of strain along its short edge (the transverse direction). At evenly spaced temperatures between 5 and 300 K, this bias was swept from -20 V to 100 V at a speed of 1 V/s, repeated multiple times to reset the strain state and accurately record any hysteretic effects.

Voltage was converted to strain using a P-doped silicon strain gauge (PiezoMetrics) glued to the underside of the piezostack and measured in tandem. Using the well known strain response of the silicon allowed for precise in-situ estimation of the strain experienced by samples. Strain on the axis transverse to the poling direction can be calculated using the well known Poisson ratio of the piezo stack \cite{Kuo2013}.

Due to the 0-crossing of the E$_g$ elastoresistivity, the strain response of any direction of resistivity will be parallel at approximately 225 K, as all strain dependence at this temperature comes from A$_{1g}$ effects. As there should be no strain-induced anisotropy at this temperature, it provides a convenient launching-point for finding the magnitude of strain experienced by the sample as a consequence of the thermal expansion of the piezoelectric on which it's mounted. A more detailed methodology is written up in the supplementary material of this paper.

We measured the off-diagonal elastoresistivity $m_{21}$ using two independent techniques: an AC method (Fig.~\ref{fig1}j) and a DC method (Fig.~\ref{Fig2}d). The two datasets are in good agreement, both exhibiting a rapid downturn below $T_{\mathrm{CDW}}$ that reaches a negative peak near $T\approx180$~K, followed by an additional slope change at lower temperatures. The AC measurement, however, shows a sharp peak at $T_{\mathrm{CDW}}$ whose origin is not yet understood. As shown in Fig.~4d, this feature is highly sensitive to the DC strain offset, suggesting that it may arise from nonlinear strain coupling to ferroaxial fluctuations. Elucidating its origin remains an open question for future work.

\subsection{Curie Weiss Fitting}

The curve in Figure \ref{fig3}C shows the Curie-Weiss-like behavior of the E$_g$ elastoresistivity of 1T-TiSe$_2$. In order to demonstrate that it follows the form $\chi_{nem} \propto (T-T^*)^{-1}$, the scipy curve\_fit package was used to find parameters best matched to the data, finding a T$^*$ equal to 172.5 K. The range of points included in the fit was determined by iterating over fit ranges to maximize the R$^2$ value of the fit while still including a statistically significant number of points. The fit shown in the paper uses points from 300 K to 180 K and has an R$^2$ of 0.9899. 

We note that the peak temperature of the nematic susceptibility observed in the DC elastoresistivity measurements ($T_{peak} \approx 165$ K, Fig. \ref{fig3}c) is slightly lower than the Weiss temperature extracted from the Curie-Weiss fit ($T^* \approx 170$ K) and the nematic features observed in the zero-strain AC trace (Fig. \ref{fig4}). We attribute this shift to the unavoidable thermal strain inherent to the DC configuration, where the sample is fully glued to the piezoelectric stack. Due to the differential thermal contraction between the sample and the stack, a finite background strain develops upon cooling. Comparing this result with the global phase diagram constructed in Fig. \ref{fig4}h, the lower peak temperature is consistent with the sample being in a slightly tensile strained state, which suppresses the nematic transition. Furthermore, the fact that $T^* > T_{peak}$ is consistent with the behavior of a three-state Potts transition in the presence of a symmetry-breaking strain field, which enhances the susceptibility relative to the zero-strain value due to the cubic term in the free energy.

\subsection{Symmetry of the elastoresistivity tensor}

The elastoresistivity tensor defined by $\rho_{ij} = m_{ijkl} \varepsilon_{kl}$ can be decomposed into irreducible representations (irreps) of the high-temperature point group $D_{3d}$ by reducing the product representation of the resistivity tensor $\rho_{ij}$ and the strain tensor $\varepsilon_{ij}$ \cite{PhysRevB.92.235147,DayRoberts2025}. The in-plane components of the resistivity transform as $\rho_{A_{1g}}=\rho_{xx}+\rho_{yy}$ and $\rho^\alpha_{E_{g}}=(\rho_{xx}-\rho_{yy},2\rho_{xy})$ where $\alpha=1,2$ labels the irreducible representation (irrep) component. The same decomposition holds for the strain tensor. Disregarding the trivial irreps, we can write the symmetry-adapted elastoresistivity for the in-plane components as $\rho^\alpha_{E_{g}} = m_{\alpha \beta} \varepsilon^\beta_{E_g}$ with $m_{11}+m_{22} \sim A_{1g}$, $m_{12}-m_{21}\sim A_{2g}$ and $(m_{11}-m_{22},m_{12}+m_{21})\sim E_g$. In particular, we note that $m_{12}$ and $m_{21}$ contain contributions from both $A_{2g}$ and $E_{g}^{(2)}$ (the second component of the $E_g$ irrep). Because the relative sign between these two contributions changes between $m_{12}$ and $m_{21}$, we can isolate them by writing the symmetric and antisymmetric combinations of these two coefficients.

Elastoresistivity can then be used to probe the symmetry of the different phase transitions in two ways. In the high-temperature phase, the diagonal components $m_{11}=m_{22}$ serve as a generalized susceptibility to probe an $E_g$ order parameter, i.e. a nematic susceptibility, which is enhanced near a nematic transition. The non-trivial irreps of $m_{\alpha \beta}$ are not allowed in the high-temperature phase, but serve directly as proxies of the order parameters with the corresponding broken symmetries below the transition: $m_{12}-m_{21}$ is a ferroaxial order parameter, signaling the breaking of all vertical mirrors and in-plane two-fold axes, while $(m_{11}-m_{22},m_{12}+m_{21})$ signals the breaking of rotational symmetry. Incidentally, $\rho^\alpha_{E_{g}}$ itself can also serve as a nematic order parameter.

\subsection{Characterization of Features in the Elastocaloric Response}

Fig. \ref{fig4}b shows a waterfall plot of the derivative of the elastocaloric coefficient of 1T-TiSe$_2$ with respect to temperature at various offset strains. The temperatures at which we see features in this derivative, marked in the figure by circular markers, correspond to various transitions in the material. These features were identified by using the scipy curve\_fit package to fit data to a sum of four Voigt peaks. The markers in Figure \ref{fig4}b each correspond to the center of one Voigt peak in the fit. 

\subsection{Elasto X-ray}
To simultaneously assess the CDW structural response to applied uniaxial stress, we performed elasto X-ray diffraction measurements following the technique established in our previous work. Single-crystal samples of 1T-TiSe$_2$ were mounted across the gap of a Razorbill CS-100 piezoelectric strain device using Stycast epoxy. Measurements were conducted at beamline 6-ID-B of the Advanced Photon Source at Argonne National Laboratory. X-rays with an energy of 11.215 keV were used to illuminate the central cross-section of the crystal where strain transmission is highly uniform. The sample and strain cell were mounted in a closed-cycle cryostat to perform isothermal strain sweeps. The applied nominal strain $\varepsilon_{nom}$ was quantified continuously during the X-ray measurements using an integrated capacitance strain gauge.

\section*{Note added}
As this work was concluded, two complementary experimental studies on 1T-TiSe$_2$ appeared. The paper Jiang \textit{et al.} \cite{Jiang2026}, of which one of us (RMF) is a co-author, independently identifies ferroaxial order using elastoresistivity, corroborating our $A_{2g}$ symmetry assignment. Lv \textit{et al.} \cite{Lv2026} report evidence for nematic fluctuations in Cu$_x$TiSe$_2$, consistent with the $E_g$ instability identified here. Our work unifies these observations within a single symmetry-breaking hierarchy.

\begin{acknowledgments}
We thank Q. Jiang for sharing her results prior to the submission of Ref. \cite{Jiang2026}. We thank T. Birol, E. Day-Roberts, Q. Jiang and I. R. Fisher for fruitful discussions. The material synthesis and the strain experiments performed at the University of Washington were primarily supported by the University of Washington Molecular Engineering Materials Center, a
U.S. National Science Foundation Materials Research
Science and Engineering Center (DMR-2308979). The work done by S.E. and J.W was supported in part by the Gordon and Betty Moore Foundation, Grant GBMF13842. M.N.G. is supported by grants RYC2021-031639-I and PID2023-153277NB-I00 funded by MCIN/AEI/10.13039/501100011033 and European Union NextGenerationEU/PRTR. F.J. acknowledges support from a 2024 Leonardo Grant for Scientific Research and Cultural Creation, BBVA Foundation. I.M. acknowledges financial support by the Swiss National Science Foundation (SNSF) via the SNSF postdoctoral Grant No. TMPFP2\_217204. This research used resources of the Advanced Photon Source, a U.S. Department of Energy (DOE) Office of Science User Facility operated for the DOE Office of Science by Argonne National Laboratory under Contract No. DE-AC02-06CH11357.  X.X. and J.-H.C. acknowledge support from the State of Washington-funded Clean Energy Institute.
\end{acknowledgments}

\bibliography{bibliography}

\appendix

\renewcommand{\thefigure}{A\arabic{figure}}
\renewcommand{\thetable}{A\arabic{table}}
\renewcommand{\theequation}{A\arabic{equation}}

\setcounter{figure}{0}
\setcounter{table}{0}
\setcounter{equation}{0}
\setcounter{section}{0}


\section{Sample Preparation and Images}

\begin{figure}[ht]
\centering
\includegraphics[width=0.4\textwidth]{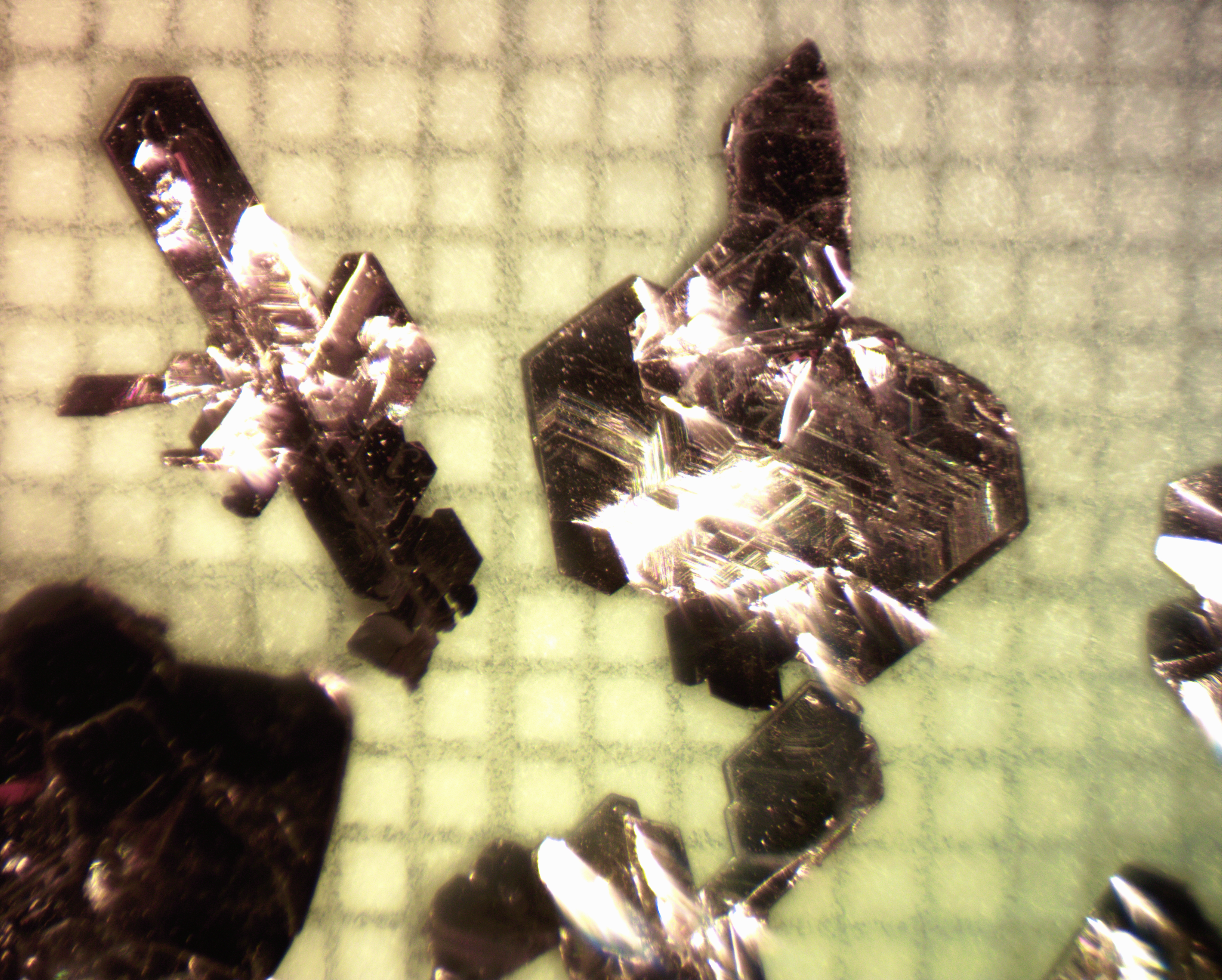}
\caption{1T-TiSe$_2$ crystals.} 
\label{sfig1}
\end{figure}

1T-TiSe$_2$ crystals grown via chemical vapor transport form in large plates with clear 120$^\circ$ facets and a slight purple sheen. Fig. \ref{sfig1} shows an image of some of these crystals.

\begin{figure}[ht]
\centering
\includegraphics[width=0.4\textwidth]{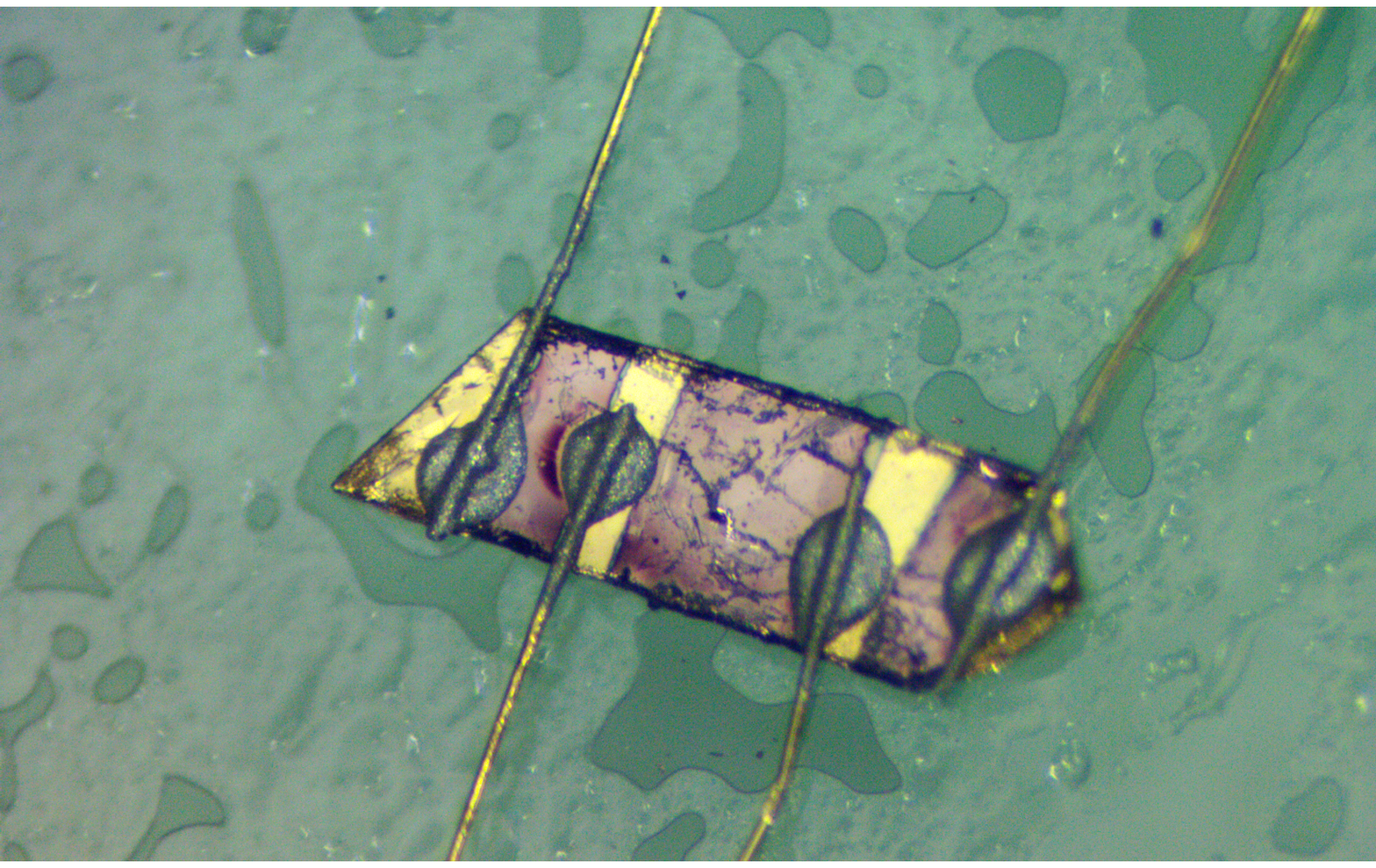}
\caption{A bar of 1T-TiSe$_2$ with gold pads and gold wire contacts.} 
\label{sfig2}
\end{figure}

These crystals are cleaved apart with a sharp scalpel blade into level $\sim$20 $\mu m$ thick plates before they are cut into different sample geometries. Before making contacts on the surface of the crystal, gold is sputtered onto the surface in the spots where silver epoxy will make contact, in order to ensure the best possible electrical connectivity between wires and sample surface. Gold wires are attached using 2-part silver epoxy. An example of a contacted sample is shown in Fig. \ref{sfig2}.

\begin{figure}[ht]
\centering
\includegraphics[width=0.4\textwidth]{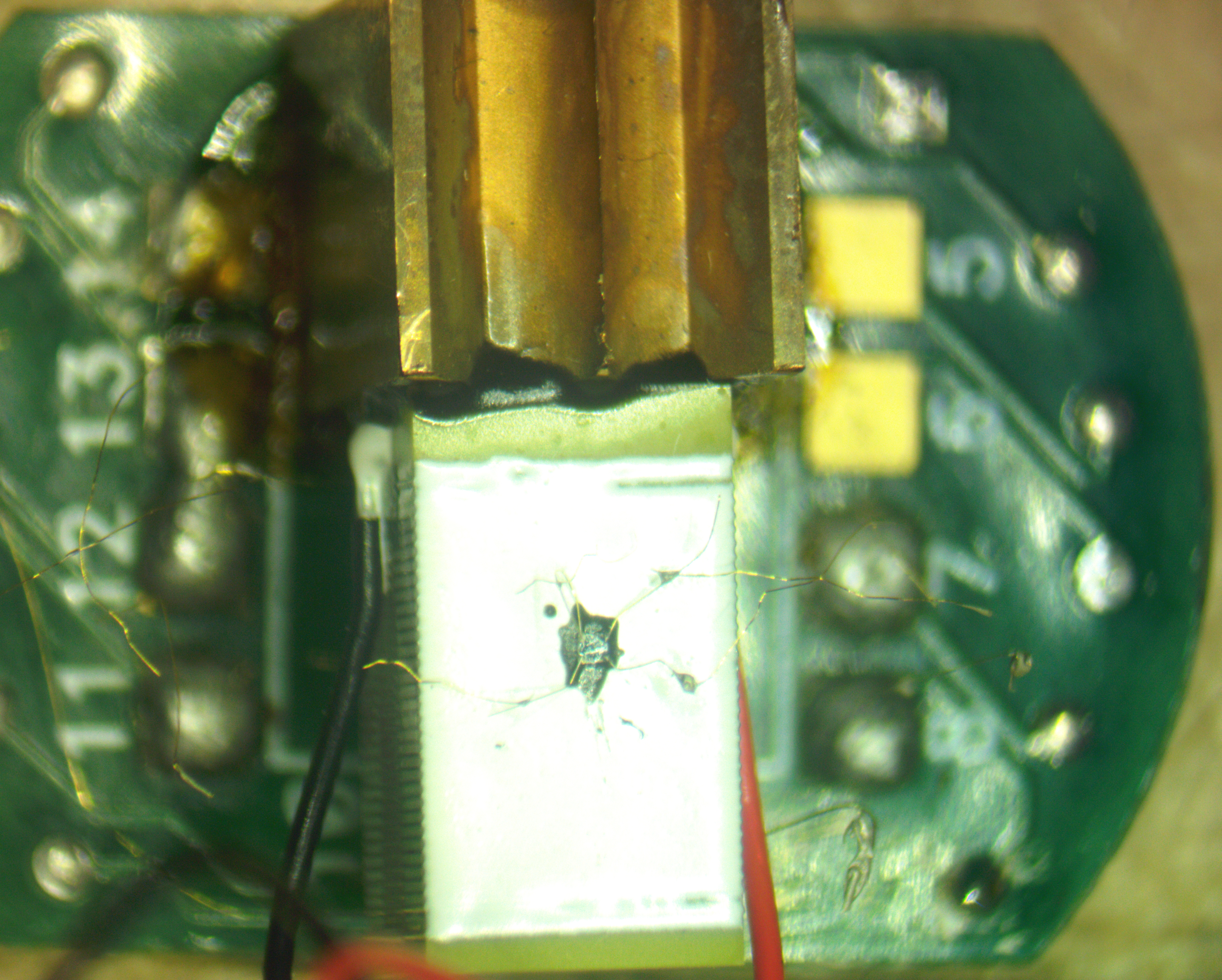}
\caption{A square of 1T-TiSe$_2$ glued to the sidewall of a piezoelectric stack.} 
\label{sfig3}
\end{figure}

Samples are mounted on strain cells or piezostacks as detailed in the methods section. An example is shown in Fig \ref{sfig3}. 

\section{Determination of Resistivity and Calculation of Neutral Strain Point}

There are several factors to consider during a measurement of symmetry-decomposed elastoresistivity. The first is accurate conversion from resistance to resistivity. For a hall bar sample geometry, this is trivial, as resistance and resistivity follow a proportional relationship, but in the modified Montgomery geometry, this relationship is slightly more complex. In this geometry, as shown in the inset of Fig. 3 in the main text, current is supplied between two corner contacts along a single side of a square sample, and voltage measured between the two other corner contacts which sit parallel to them. The resulting nonuniform electrical current density leads to the following relationships, as calculated in C.A.M dos Santos' paper proposing this approach \cite{Santos2011}:

\begin{equation}
    \rho_{xx} = \frac{\pi}{8}\left(\frac{L'_z L'_y}{L'_x}\right)\left(\frac{L_x}{L_y}\right)R_{xx}\sinh\left(\frac{\pi L_y}{L_x}\right)
\end{equation}

\begin{equation}
    \rho_{yy} = \frac{\pi}{8}\left(\frac{L'_z L'_x}{L'_y}\right)\left(\frac{L_y}{L_x}\right)R_{yy}\sinh\left(\frac{\pi L_x}{L_y}\right)
\end{equation}

Here $L'_{x,y,z}$ refer to actual sample dimensions and $L_{x,y}$ to calculated values. Many similar elastoresistance experiments use the approximation:

\begin{equation}
    \frac{L_y}{L_{x}} \approx \frac{1}{2} \left[\frac{1}{\pi}\ln\left(\frac{R_{yy}}{R_{xx}}\right)  +\sqrt{\left(\frac{1}{\pi}\ln\left(\frac{R_{yy}}{R_{xx}}\right)\right)^2 + 4}\right]
\end{equation}

but this is only valid in the limit that the resistance ratio $R_{yy}/R_{xx} > 1$. Due to the extremely large anisotropic strain response of 1T-TiSe$_2$ demonstrated in this paper, this condition is often not uniformly met across the whole strain range. Instead, we use the relationship:

\begin{equation}
    \frac{R_{yy}}{R_{xx}} \approx \frac{\sinh\left[\frac{\pi L_y}{L_x}\right]}{\sinh\left[\frac{\pi L_x}{L_y}\right]}
\end{equation}

We solve this equation numerically (introducing minimal error on the order $\sim 10^{-5}$\%) in order to calculate $L_y/L_x$ as a function of $R_{xx}/R_{yy}$. Additionally, to allay additional uncertainty introduced by manual measurement of the physical parameters $L'_x$ and $L'_y$, we can use known constraints on the resistivity tensor in order to make more precise approximations of their values. 

\begin{figure*}
\centering
\includegraphics[width=0.9\textwidth]{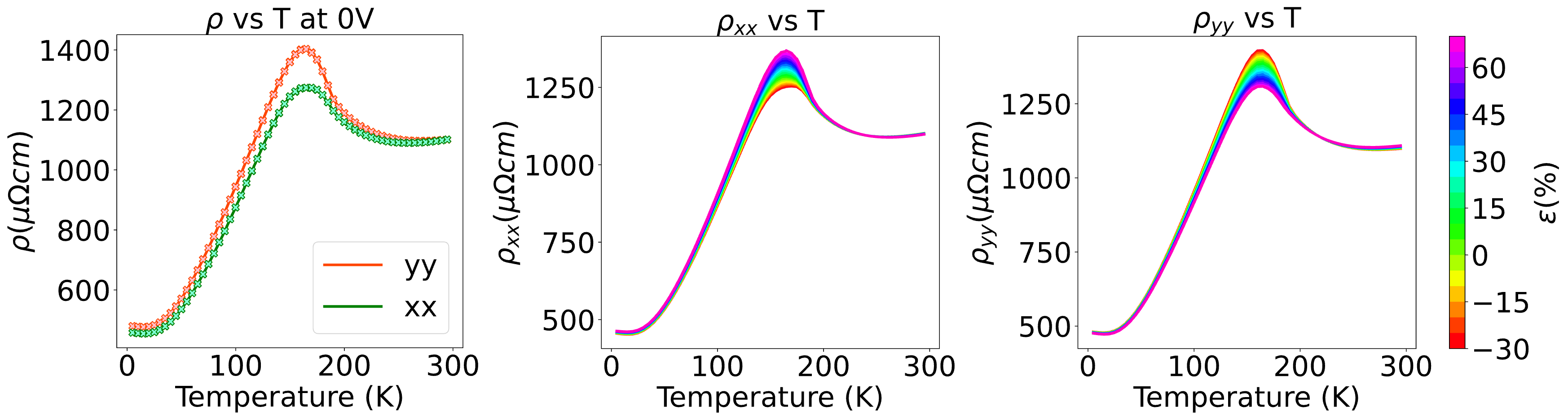}
\caption{Resistivity of a Montgomery square mounted on a piezostack at different voltages. The resistivity has been calculated using the condition $\rho_{xx}(300 K) = \rho_{yy}(300K)$} 
\label{sfig4}
\end{figure*}

There are two instinctive ways to do this. First is the assertion that $\rho_{xx}$ and $\rho_{yy}$ must be equivalent in the absence of any known symmetry breaking order at room temperature (300 K). This gives us the condition that $L'_y/L'_x = L_y/L_x$ at room temperature on a sample mounted on a piezo stack receiving no voltage bias, assuming that no strain has been applied to the sample by the mounting process. Using data from the experiment depicted in Fig. 3 of the main text as an example, this produces the resistivity shown in Fig. \ref{sfig4}.

The challenge to this approach becomes apparent when the neutral strain point is calculated. This is traditionally done using a similar principle: above any known symmetry-breaking phase transition, an unstrained crystal should have no anisotropy between $\rho_{xx}$ and $\rho_{yy}$ \cite{Liu2024}. Therefore, while sweeping piezo bias voltage at a fixed temperature, the point at which the two quantities are equivalent is the point at which applied anisotropic strain has canceled out strains caused by the opposing thermal expansions of the piezo stack and crystal. 

\begin{figure*}
\centering
\includegraphics[width=0.9\textwidth]{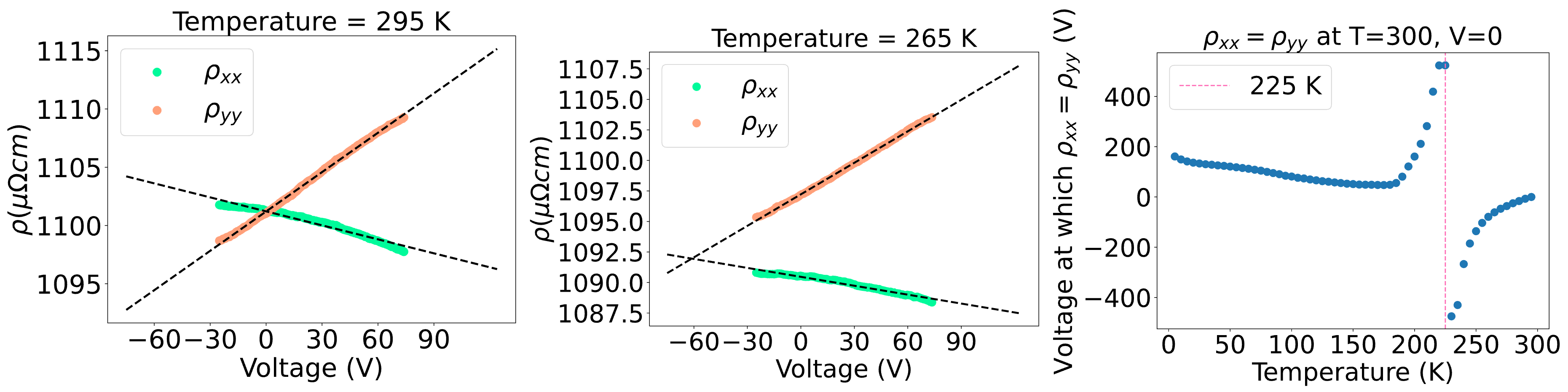}
\caption{Calculation of the neutral strain point by identifying voltages at which $\rho_{xx}= \rho_{yy}$. Two examples of fixed-temperature voltage sweeps are shown, one where the resistivities in each direction cross within the experiment's strain range, and one where the neutral strain point is extrapolated via line fit. The rightmost plot shows how these "crossing points" develop with temperature.} 
\label{sfig5}
\end{figure*}

Fig. \ref{sfig5} shows the results of doing this calculation when resistivity has been calculated under the condition $\rho_{xx}(300 K) = \rho_{yy}(300K)$. For temperatures where $\rho_{xx}$ and $\rho_{yy}$ never reach equivalent values, a linear fit is used to extrapolate the applied voltage where they would meet.

\begin{figure}
\centering
\includegraphics[width=0.4\textwidth]{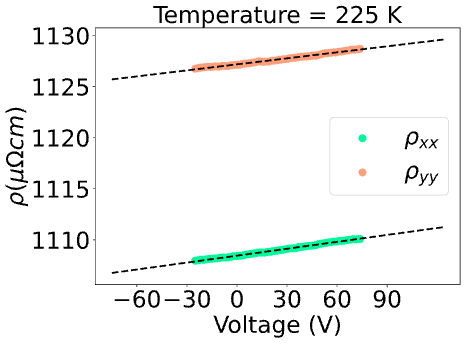}
\caption{The parallel strain responses of $\rho_{xx}$ and $\rho_{yy}$ at 225 K.} 
\label{sfig6}
\end{figure}

The large divergence seen in Fig. \ref{sfig5} is clearly artificial, as neither the piezo stack or the TiSe$_2$ is known to have any divergence in their thermal expansion coefficients\cite{Kuo2013, Caille1983}, and the temperature around which the divergence is centered is well above the charge density wave onset temperature. Referring to the anisotropic elastoresistivity seen in Fig. 3, it becomes clear what the source is. At 225 K, exactly where this divergence is seen, the $E_g$ portion of elastoresistivity goes to 0, meaning that the only contribution to the strain response should come from $A_{1g}$ effects at this temperature. Indeed, as seen in Fig. \ref{sfig6}, $\rho_{yy}(V)$ and $\rho_{xx}(V)$ are nearly exactly parallel.

\begin{figure*}
\centering
\includegraphics[width=0.9\textwidth]{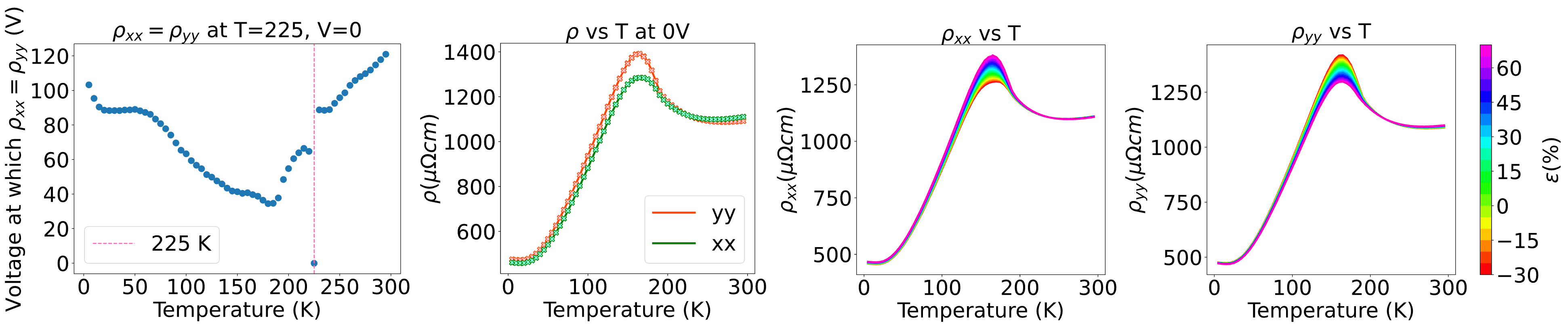}
\caption{Calculation of the neutral strain point by identifying voltages at which $\rho_{xx}= \rho_{yy}$ as well as resistivity of the Montgomery square mounted on a piezostack at different voltages. The resistivity has been calculated using the condition $\rho_{xx}(T = 225 K, V=0V) = \rho_{yy}(T=225K,V=0V)$.} 
\label{sfig7}
\end{figure*}

This leads us to a new approach in our approximation of the geometric parameters used in our resistivity calculation. Instead of holding true the condition that $\rho_{xx}(300 K) = \rho_{yy}(300K)$, it seems clear that we may get a more precise estimation by the constraint that $\rho_{xx}(225 K) = \rho_{yy}(225K)$. We can no longer assume that this condition must be met at 0 bias voltage due to the presence of nonzero thermal strain, but the calculations done under the conditions $\rho_{xx}(T = 225 K, V=0) = \rho_{yy}(T=225K,V=0)$ are shown for completeness in Fig. \ref{sfig7}.

\begin{figure*}
\centering
\includegraphics[width=0.9\textwidth]{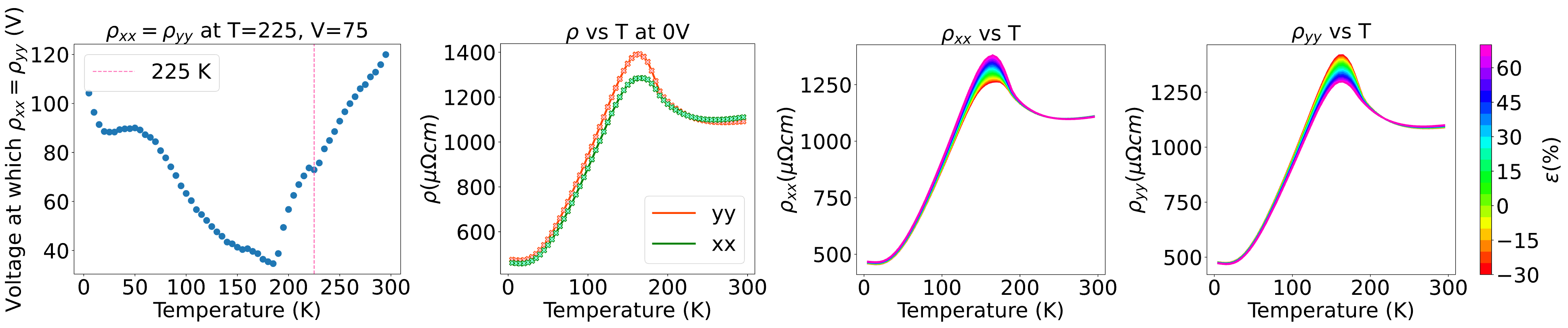}
\caption{Calculation of the neutral strain point by identifying voltages at which $\rho_{xx}= \rho_{yy}$ as well as resistivity of the Montgomery square mounted on a piezostack at different voltages. The resistivity has been calculated using the condition $\rho_{xx}(T = 225 K, V=75 V) = \rho_{yy}(T=225K,V=75V)$.} 
\label{sfig8}
\end{figure*}

Now that the divergence has largely gone, all that remains is to use the surrounding, non-parallel, points to extrapolate the true voltage at which we should expect $\rho_{xx}$ and $\rho_{yy}$ to be equivalent. Our final result is shown in Fig. \ref{sfig8}.

\section{Identification of CDW Onset via Second Derivative of Resistivity}

A peak in the second derivative of resistivity with respect to temperature, marking the onset of the anomalous resistivity hump associated with the CDW state, has been noted in many previous works to empirically be a good indicator of the transition temperature into the CDW state \cite{Moulding2022, Watson2019, Campbell2019, Li2022, Wang2025}. In this section, we present the resistivity of each sample used to take the data presented in this work and its second derivative. Data has been interpolated using the python package Scipy.interpolate in order to achieve smooth derivatives. 

\begin{figure}[ht]
\centering
\includegraphics[width=0.5\textwidth]{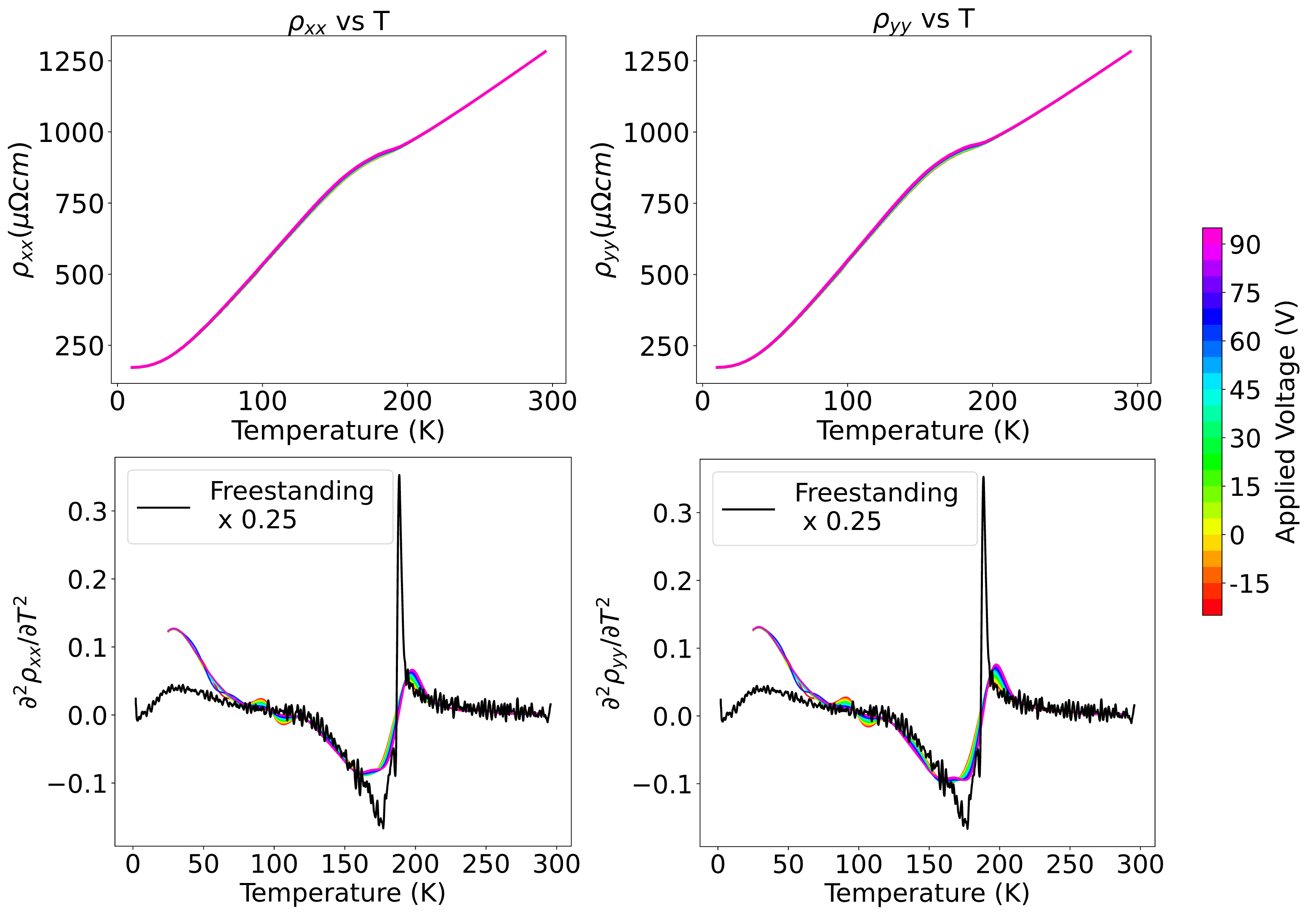}
\caption{(Top) Resistivity of a Montgomery square of 1T-TiSe$_2$ mounted at 45 degrees on a piezostack, at different piezo voltage biases. (Bottom) The second derivative with respect to temperature of the resistivities shown above, plotted alongside the second derivative of the sample's resistivity prior to gluing it onto the strain cell. Since the second derivative of the freestanding sample has a much sharper peak, its value has been scaled by a factor of 0.25 to better show both curves together.} 
\label{sfig10}
\end{figure}

\begin{figure}[ht]
\centering
\includegraphics[width=0.5\textwidth]{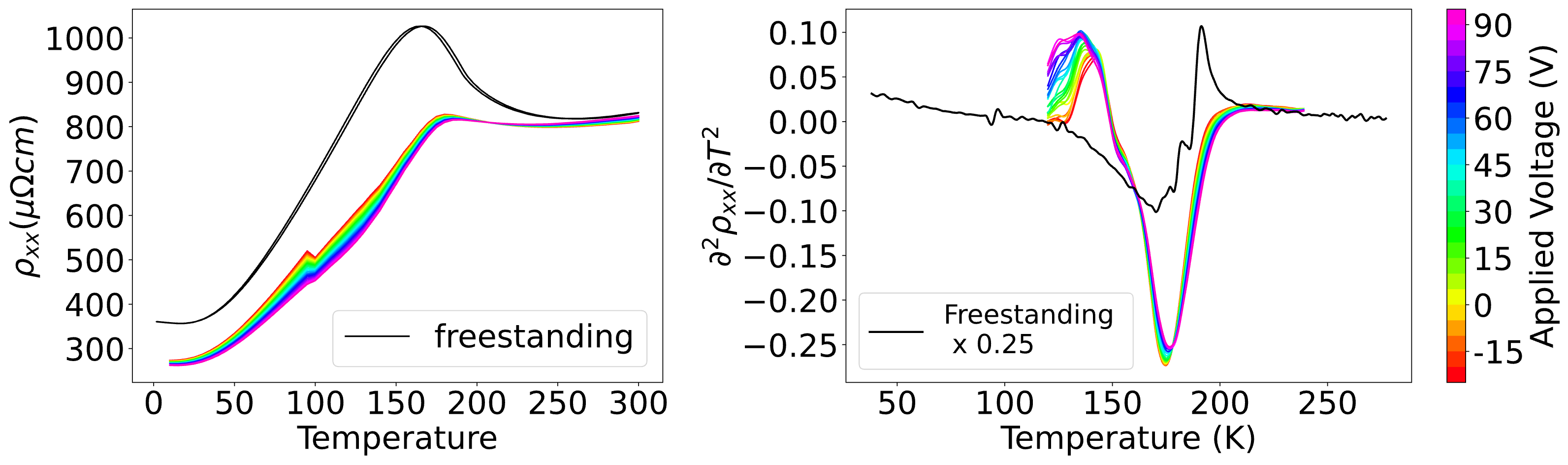}
\caption{(Left) Resistivity of a Hall bar of 1T-TiSe$_2$ mounted on a piezostack parallel to the poling direction, at different piezo voltage biases. (Right) The second derivative with respect to temperature of the resistivities at each voltage bias, plotted alongside the second derivative of the sample's resistivity prior to gluing it onto the strain cell. Since the second derivative of the freestanding sample has a much sharper peak, its value has been scaled by a factor of 0.25 to better show both curves together.} 
\label{sfig11}
\end{figure}

\begin{figure}
\centering
\includegraphics[width=0.5\textwidth]{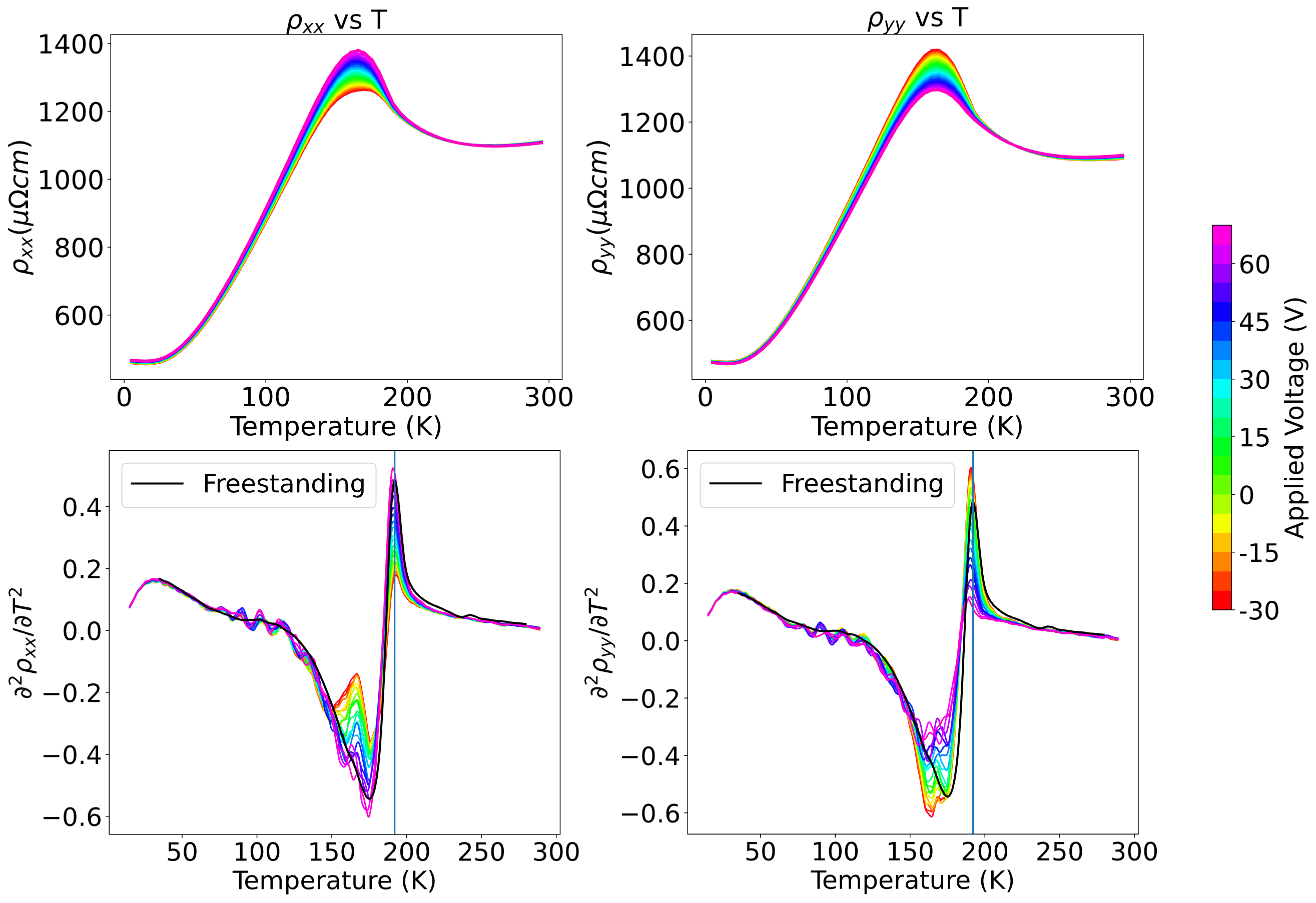}
\caption{(Top) Resistivity of a Montgomery square of 1T-TiSe$_2$ mounted on a piezostack parallel to the poling direction, at different piezo voltage biases. (Bottom) The second derivative with respect to temperature of the resistivities shown above, plotted alongside the second derivative of the sample's resistivity prior to gluing it onto the strain cell.} 
\label{sfig12}
\end{figure}

Fig. 2 of the main text includes two samples with different geometries. Fig. \ref{sfig10} corresponds to the square sample used to calculate $m_{12}$. Fig. \ref{sfig11} corresponds to the Hall bar sample used to calculate $m_{21}$. Fig 3 of the main text includes data from only one sample, whose resistivities and second derivatives are shown in Fig. \ref{sfig12}.

As discussed in the main text, the DC elastoresistivity (Fig. 2) is unable to resolve a $\approx 7$~K gap between the primary CDW and ferroaxial transitions. This apparent merging is likely a natural consequence of the DC experimental configuration. Because the transitions are separated by only a few Kelvin, critical fluctuations of the two orders heavily overlap in transport. Furthermore, the samples are fully glued to the piezoelectric stack for DC measurements, introducing a finite background thermal strain upon cooling that inherently broadens the transition boundaries.

\begin{figure*}
\centering
\includegraphics[width=0.9\textwidth]{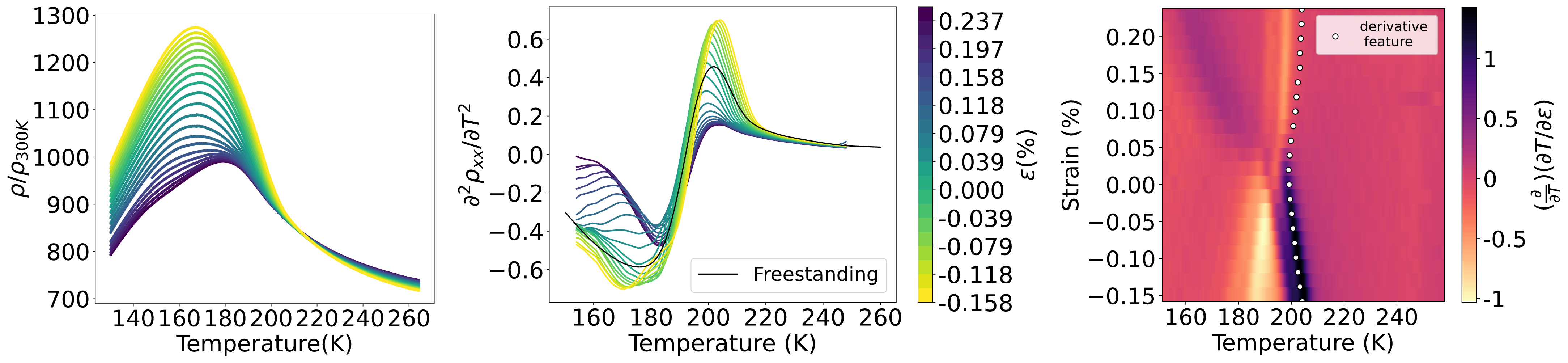}
\caption{(Left) Normalized resistivity ($\rho_{xx}/\rho_{xx}(T = 300 K, \varepsilon = 0)$) of 1T-TiSe$_2$ on a uniaxial strain cell at different DC offset strains. (Middle) Second derivative with respect to temperature of each resistivity shown to the left, plotted alongside the second derivative of the sample's resistivity prior to gluing it onto the strain cell. (Right) CDW onset temperatures at each strain gleaned from the second derivatives of resistivity plotted atop a colormap of the first derivative with respect to temperature of the elastocaloric coefficient ($(\frac{\partial}{\partial T})(\partial T/\partial\epsilon)$). The two align in shape remarkably well.} 
\label{sfig13}
\end{figure*}

Figs. 1 and 4 use data taken simultaneously on the same sample. Here the second derivative of resistivity is extremely informative, as it allows us to make informed conclusions about which feature in the material's elastocaloric response corresponds to the CDW transition. The correspondence between the two quantities is shown in Fig. \ref{sfig13}.

\begin{figure}[ht]
\centering
\includegraphics[width=0.5\textwidth]{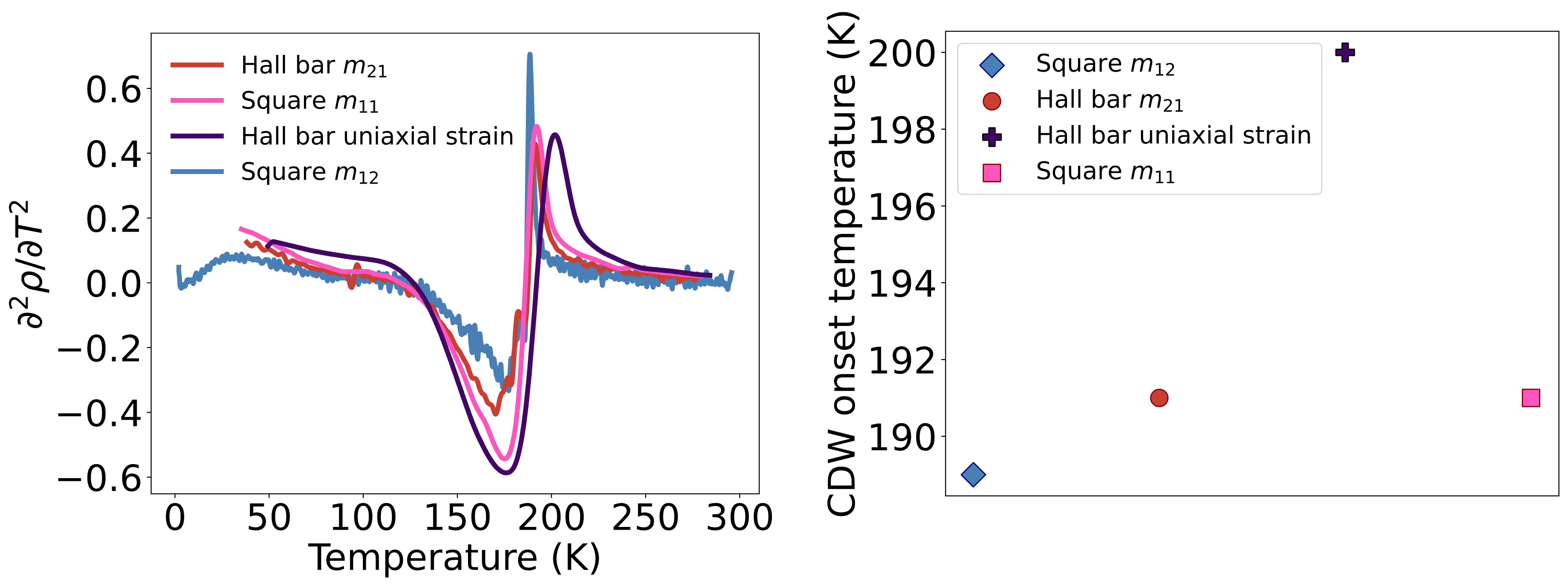}
\caption{Second derivatives with respect to temperature of the resistivity of each sample used in the main paper. (Right) The CDW onset temperature gleaned from each of these derivatives.} 
\label{sfig14}
\end{figure}

Finally, in order to show that all the samples used to draw conclusions in this paper are of comparable character, we plot the second derivative of resistivity for each sample side by side in Fig. \ref{sfig14}.

\section{Structural evidence for the strain-induced redistribution of CDW intensities}

To provide direct structural confirmation of this strain-induced redistribution, we performed X-ray diffraction measurements of the superlattice peaks at 150~K as a function of uniaxial strain. We observe a strong modulation of the CDW intensities, mirroring the symmetry-breaking mechanism implied by our thermodynamic data. Specifically, under compressive strain, the intensity of the $Q$-vector aligned with the strain axis is strongly enhanced, modulating by a factor of roughly seven over the applied strain range, while the transverse symmetry-equivalent $Q$-vectors exhibit the opposite strain dependence. 

Although the accessible strain in this specific X-ray configuration did not allow us to fully cross the phase boundary into a pure $1Q$ or $2Q$ state, this pronounced asymmetric response provides crucial structural validation for our elastocaloric phase diagram. Furthermore, it helps establish the trajectory of the highly-strained states: compressive strain explicitly favors the single wavevector aligned with the stress axis, indicating an approach toward the $1Q$ phase, whereas tensile strain promotes the transverse wavevectors, consistent with the development of the $2Q$ phase.

\begin{figure}[ht]
\centering
\includegraphics[width=0.5\textwidth]{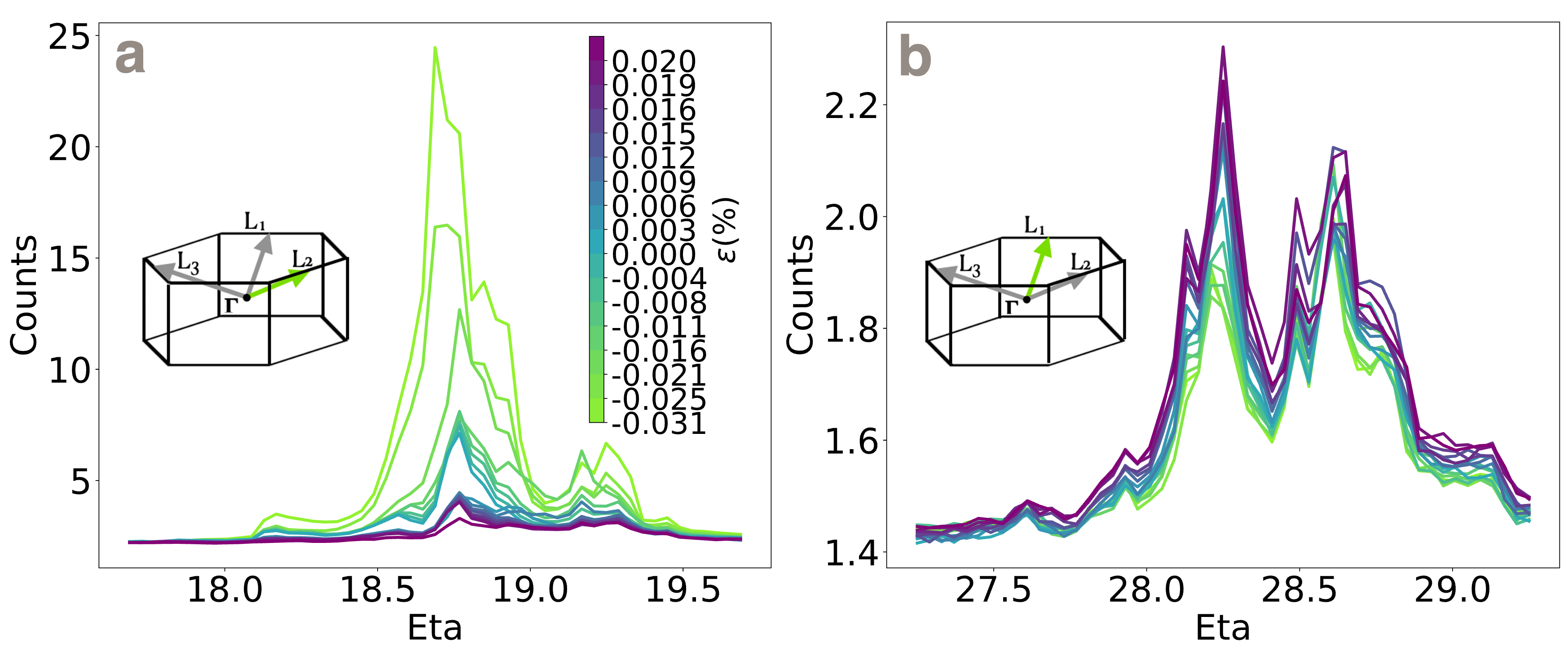}
\caption{\textbf{Strain-induced redistribution of CDW intensities.} X-ray diffraction measurements of the CDW superlattice peaks taken at $T = 150$~K under applied uniaxial strain. The color bar indicates the applied strain $\varepsilon$, ranging from compressive (green) to tensile (purple). 
\textbf{a.} Diffraction profiles of the superlattice peak associated with the wavevector aligned with the applied strain axis ($L_2$, highlighted in the inset). Compressive strain strongly favors this wavevector, enhancing its intensity by roughly a factor of seven over the measured strain range. 
\textbf{b.} Diffraction profiles for a transverse, symmetry-equivalent wavevector ($L_1$, highlighted in the inset). This peak exhibits the opposite response, being suppressed by compressive strain and enhanced by tensile strain. This highly asymmetric, strain-dependent intensity modulation provides direct structural evidence for the lifting of the $3Q$ rotational degeneracy, consistent with the system approaching a $1Q$ state under compression and a $2Q$ state under tension.} 
\label{exfig2}
\end{figure}

\section{Landau Theory of the Elastocaloric Response}

To understand the thermodynamic signatures of the elastocaloric effect (ECE) across the complex phase diagram of 1T-TiSe$_2$, we construct a Landau free energy model. The primary charge density wave (CDW) is described by 
a three-component order parameter $\vec \psi=(\psi_1, \psi_2, \psi_3)$ that transforms as the $L^-_1$ irrep of the point group, with each component associated to an equivalent $\mathbf{Q}_n=\Gamma$-$L_n$ wavevector,
\begin{align} 
\mathbf{Q}_1 &= \frac{1}{2}(\mathbf{a}^* + \mathbf{c}^*) = \left(0, \frac{2\pi}{\sqrt{3}}, \pi\right) \\ 
\mathbf{Q}_2 &= \frac{1}{2}(\mathbf{b}^* + \mathbf{c}^*) = \left(-\frac{\sqrt{3}}{2}\frac{2\pi}{\sqrt{3}}, -\frac{1}{2}\frac{2\pi}{\sqrt{3}}, \pi\right) \\ 
\mathbf{Q}_3 &= \frac{1}{2}(-\mathbf{a}^* + \mathbf{b}^* + \mathbf{c}^*) = \left(\frac{\sqrt{3}}{2}\frac{2\pi}{\sqrt{3}}, -\frac{1}{2}\frac{2\pi}{\sqrt{3}}, \pi\right) 
\end{align}

and with the hexagonal reciprocal lattice basis $\mathbf{a}^*$, $\mathbf{b}^*$, and $\mathbf{c}^*$. 
Ignoring the secondary CDW mode and phase fluctuations, the unperturbed free energy expansion up to the quartic term reads
\begin{equation}
    \mathcal{F}_0=a(T-T_\mathrm{CDW})\vec{\psi}^2+b\vec{\psi}^4+c(\psi_1^4+\psi_2^4+\psi_3^4)\label{eq:F0}
\end{equation}
where $T_\mathrm{CDW}$ is the bare transition temperature, $\vec\psi^2=\sum_n  \psi_n^2$ for $n=\{1,2,3\}$ and $\vec\psi^4=(\vec\psi^2)^2$. 
Here, $a>0$, $b>0$, and $b+c>0$ are required for the free energy to remain bounded from below. When $T<T_\mathrm{CDW}$, the system condenses into a state with a finite order parameter, whose internal structure is determined by the sign of $c$. For $c<0$ we have a $1Q$ state, with $\vec\psi \sim (1,0,0)$, and for $c>0$ the $C_3$-symmetric $3Q$ state, with $\vec\psi \sim (1,1,1)$. Since the last one is the experimentally relevant state below the CDW transition, we will choose $c>0$. 

\begin{figure*}
\centering
\includegraphics[width=0.9\textwidth]{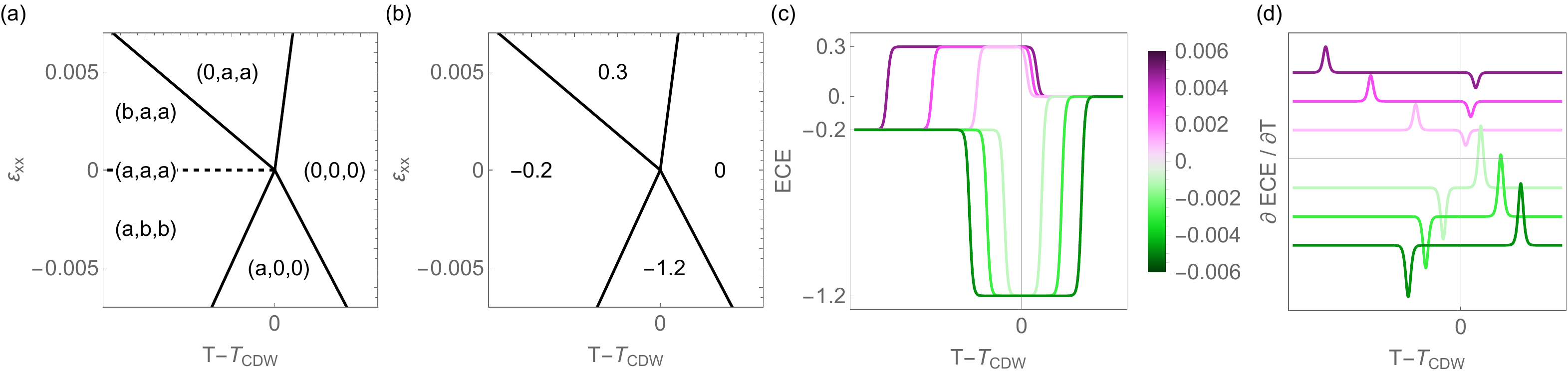}
\caption{\textbf{a}. Theoretical prediction of the phase diagram of 1T-TiSe$_2$ in strain-temperature space for the case of a single CDW instability (Eqs.\eqref{eq:T1Q}-\eqref{eq:T2Q} and Eqs.\eqref{eq:psi1-3Q}-\eqref{eq:psi2-3Q}). \textbf{b}. Theoretical value of the elastocaloric coefficient in each region of the phase diagram (Eqs.\eqref{eq:ECE-1Q},\eqref{eq:ECE-2Q},\eqref{eq:ECE-3Q}). \textbf{c}. The elastocaloric coefficient for different DC offset strains. \textbf{d}. The derivative with respect to temperature of the elastocaloric coefficient shown (c). The temperature versus strain 2D map is shown in Fig. 4(f) of the main text, with $\varepsilon_0=a T_\mathrm{CDW}/\lambda_E$. In all panels the parameters are $a=1$, $T_\mathrm{CDW}=1$, $b=0.25$, $c=0.5$, $\lambda_E=1$ and $\lambda_A=0.2$.} 
\label{sfig_t}
\end{figure*}

Applying a uniaxial strain $\varepsilon$ along the $\mathbf{Q}_1$ direction introduces an explicit symmetry breaking. The strain decomposes into a totally symmetric $A_{1g}$ component, $\varepsilon_A=\varepsilon_{xx}+\varepsilon_{yy}$, and a symmetry-breaking $E_g$ component, $\varepsilon_E=\varepsilon_{xx}-\varepsilon_{yy}$. To lowest order, the strain coupling to the order parameter is:
\begin{equation}
    \mathcal{F}_\mathrm{strain} = \lambda_A \varepsilon_A \vec{\psi}^2 + \lambda_E \varepsilon_E \left(\psi_1^2 - \frac{1}{2}\psi_2^2 - \frac{1}{2}\psi_3^2\right)\label{eq:Feps}
\end{equation}
Here, $\lambda_A$ couples volume changes to the total amplitude of the CDW, while $\lambda_E$ acts as a nematic coupling that redistributes the CDW amplitude between the constituent wave-vectors. Gathering the quadratic terms yields the split transition temperatures:
\begin{align}
T_{c,1Q} &= T_\mathrm{CDW} -\frac{\lambda_A}{a}\varepsilon_A-\frac{\lambda_E}{a}\varepsilon_E \label{eq:T1Q} 
\\
T_{c,2Q} &= T_\mathrm{CDW} -\frac{\lambda_A}{a}\varepsilon_A+\frac{\lambda_E}{2a}\varepsilon_E \label{eq:T2Q} 
\end{align}
shown in Fig.\ref{sfig_t}(a).

Under adiabatic conditions, the elastocaloric coefficient can be expressed in terms of derivatives of the entropy with respect to strain and temperature
\begin{equation}
    \left(\frac{\partial T}{\partial \varepsilon}\right)_S=-\frac{\left(\frac{\partial S}{\partial \varepsilon}\right)_T}{\left(\frac{\partial S}{\partial T}\right)_\epsilon}=  -\frac{T}{C_\varepsilon} \left(\frac{\partial S}{\partial \varepsilon}\right)_T,
\end{equation}
where $C_\varepsilon$ is the specific heat under constant strain.
Within the minimal model for the CDW order parameter, the entropy 
is given by
$S= -\frac{\partial \mathcal{F}}{\partial T}= -a\vec{\psi}^2$, and the ECE signal is proportional to the strain derivative of the total squared order parameter: 
\begin{equation}
    \left(\frac{\partial T}{\partial \varepsilon}\right)_S=-\frac{\partial_\varepsilon \vec{\psi}^2}{\partial_T \vec{\psi}^2}.
\end{equation}
where $\varepsilon\equiv \varepsilon_{xx}=(\varepsilon_A + \varepsilon_E)/2$ in the experiment. We can now calculate the ECE coefficient across the distinct phase boundaries:

\noindent \textbf{1. The $1Q$ Phase (Compressive Strain, $\varepsilon < 0$):}
Assuming $\lambda_E > 0$, compressive strain stabilizes $\psi_1$ first. Minimizing the free energy with respect to $\psi_1$ yields 
\begin{equation}
\begin{split}
    \psi_1^2 =\vec{\psi}^2_{1Q}= -\frac{a(T-T_\mathrm{CDW})+\lambda_A\varepsilon_A+\lambda_E\varepsilon_E}{2b+2c}  \\\\=-\frac{a(T-T_{c,1Q})}{2b+2c}
\end{split}
\end{equation}
The ECE signal upon entering the $1Q$ state is therefore a constant strain-independent step:
\begin{align}
    \partial_{\varepsilon_A} \vec{\psi}^2_{1Q}&=-\frac{\lambda_A}{2b+2c}=\frac{\lambda_A}{\lambda_E}\partial_{\varepsilon_E} \vec{\psi}^2_{1Q}\\
    \partial_T\vec{\psi}^2_{1Q}&=-\frac{C_{\varepsilon,1Q}}{aT}=-\frac{a}{2b+2c}\\
\left(\frac{\partial T}{\partial \varepsilon}\right)_{S,1Q}&=-\frac{\lambda_A+\lambda_E}{a}\label{eq:ECE-1Q}
\end{align}
where we used the fact that $  \partial_{\varepsilon} \vec{\psi}^2_{1Q}= \partial_{\varepsilon_A} \vec{\psi}^2_{1Q}+\partial_{\varepsilon_E} \vec{\psi}^2_{1Q} $. 

\noindent \textbf{2. The $2Q$ Phase (Tensile Strain, $\varepsilon > 0$):}
Under tensile strain, the transverse modes $\psi_2 = \psi_3 \equiv \psi$ condense first. Minimizing the free energy and summing over the two components yields,
\begin{equation}
    \vec{\psi}^2_{2Q}=-\frac{a(T-T_\mathrm{CDW})+\lambda_A\varepsilon_A-\lambda_E\varepsilon_E/2}{2b+c}=-\frac{a(T-T_{c,2Q})}{2b+c}
\end{equation}
The ECE signal upon entering the $2Q$ state is then:
\begin{align}
\partial_{\varepsilon_A} \vec{\psi}^2_{2Q}&=-\frac{\lambda_A}{2b+c}=-\frac{\lambda_A}{\lambda_E/2}\partial_{\varepsilon_E}\vec{\psi}^2_{2Q}\\
    \partial_T\vec{\psi}^2_{2Q}&=-\frac{C_{\varepsilon,2Q}}{aT}=-\frac{a}{2b+c}\\
\left(\frac{\partial T}{\partial \varepsilon}\right)_{S,2Q}&=-\frac{\lambda_A-\lambda_E/2}{a}\label{eq:ECE-2Q}
\end{align}

\noindent \textbf{3. The $3Q$ Phase (Low Temperature):}
Once all three components are condensed at lower temperatures, the total squared order parameter 
solution is now independent of the $E_g$ component of the strain, since
\begin{align}
    \psi_1^2 &=-\frac{a(T-T_\mathrm{CDW})+3\lambda_A\varepsilon_A}{6b+2c}-\frac{\lambda_E\varepsilon_E}{2c} \label{eq:psi1-3Q}\\
    \psi_2^2 &=\psi_3^2=-\frac{a(T-T_\mathrm{CDW})+3\lambda_A\varepsilon_A}{6b+2c}+\frac{\lambda_E\varepsilon_E}{4c}\label{eq:psi2-3Q}
\end{align}
and hence,
\begin{equation}
    \vec{\psi}^2_{3Q}=-\frac{3a(T-T_\mathrm{CDW})+3\lambda_A\varepsilon_A}{6b+2c}
\end{equation}
Note that the transition temperature between the $1Q$-$3Q$ states and the $2Q$-$3Q$ states can be obtained by the conditions $\psi_{1}^2=0$ and $\psi_{2}^2=0$, respectively (Fig.\ref{sfig_t}(a)).
The derivative yields the bulk ECE signal:
\begin{align}
\partial_{\varepsilon_A} \vec{\psi}^2_{3Q}&=-\frac{3\lambda_A}{6b+2c};\quad \partial_{\varepsilon_E}\vec{\psi}^2_{3Q}=0\\
    \partial_T\vec{\psi}^2_{3Q}&=-\frac{C_{\varepsilon,3Q}}{aT}=-\frac{3a}{6b+2c}\\
\left(\frac{\partial T}{\partial \varepsilon}\right)_{S,3Q} &=-\frac{\lambda_A}{a} \label{eq:ECE-3Q}
\end{align}

This derivation establishes three critical features observed in the experiment. First, the mean-field theory predicts discrete jumps in the ECE at the phase boundaries, whose magnitudes depend explicitly on the coupling constants. The smearing of these steps in the experimental data likely arises from the strong $A_{1g}$ dependence of the transition temperature combined with thermal fluctuations. Second, the magnitude of the ECE step in the $1Q$ phase is mathematically distinct from the $2Q$ phase, naturally explaining the observed asymmetry between the compressive and tensile regimes (Fig.\ref{sfig_t}(b)-(c)). 
Finally, this coupling explains the absence of a large ECE signal at the zero-strain $165$~K nematic transition: the spontaneous redistribution of wave-vector intensities (the diverging $E_g$ susceptibility) largely preserves the total volume of the order parameter $\sum \psi_i^2$, rendering the purely symmetric ECE probe less sensitive to it. As explained in the main text, this model however misses a crucial experimental feature: the second phase transition close to the CDW transition, where ferroaxial order emerges. This is not unexpected, since the model with a single CDW instability does not support ferroaxial order in either the $1Q$, $2Q$, or $3Q$ phase.

\subsection*{Asymmetry of phase boundaries under compressive and tensile strain}

The experimental phase diagram (Fig.~4g of the main text) exhibits a 
striking asymmetry: the compressive side displays two distinct 
lower-temperature phase boundaries, while the tensile side displays only one. Surprisingly, the $L_1^-$ model [Eqs.\eqref{eq:F0}-\eqref{eq:Feps}] predicts only the $1Q$ to $3Q$ transition for any value of $ b,c,$ but in the data we observe an intermediate phase. From Eq.\eqref{eq:Feps} (the coupling to strain), it is clear that in the $1Q$ phase, the remaining vertical mirror symmetry requires $T_c$ for $\psi_2$ and $\psi_3$ to be the same, so when $T$ crosses this $T_c$ we get a direct transition to the $(a,b,b)$ state. The existence of an intermediate transition can thus be understood as additional evidence for the ferroaxial phase, which breaks the mirror symmetry and allows different $T_c$ for the remaining two components. Therefore, we ascribe the intermediate phase to a $2Q $ state $ (a,b,0) $ enabled by the ferroaxial order.

\subsection*{Symmetry analysis of the secondary CDW mode}

As explained in the main text, none of the $1Q$, $2Q$, or $3Q$  configurations of the $L_1^-$ CDW order parameter break the vertical mirror symmetries, and hence they do not display ferroaxial order. Consequently, the emergence of ferroaxial order requires a secondary order parameter belonging to a distinct irreducible representation. The symmetry of this secondary order parameter is strictly constrained: in the zero-strain limit, the threefold rotational symmetry is preserved right below the ferroaxial transition. Group theoretical analysis reveals that a secondary CDW order parameter transforming as $L_2^-$, at the $(1,1,1)$ configuration, $M_2^+$, at the $(1,1,1)$ configuration, or $A_2^-$ satisfies this requirement. Of course, a pure ferroaxial mode with $\Gamma_2^+$ symmetry by definition would also give rise to ferroaxial order. The difference is that, in the former cases, ferroaxial order is due to the coexistence between CDW orders that alone preserve ferroaxial symmetry.
While determining the microscopic mechanism responsible for any of these possibilities is beyond the scope of this paper, first-principles calculations consistently reveal a landscape of multiple competing CDW instabilities with nearly degenerate ground-state energies \cite{Wickramaratne2022,Wegner2020,Kim2024a,Mun25}, thus supporting the proposed scenario of two coexisting CDW orders.


\end{document}